\pdfoutput=1
\documentclass[12pt]{article}
\usepackage[a4paper,margin=2.5cm]{geometry}
\usepackage{amsmath,amssymb}
\usepackage{booktabs}
\usepackage{setspace}
\usepackage{graphicx}
\usepackage{caption}
\usepackage[flushleft]{threeparttable}
\usepackage{natbib}
\usepackage{rotating}
\usepackage[section]{placeins}
\usepackage{etoolbox}
\usepackage[hidelinks,hyperfootnotes=false]{hyperref}
\setcitestyle{aysep={}}

\AtBeginEnvironment{tabular}{\setstretch{1.0}}
\AtBeginEnvironment{tablenotes}{\setstretch{1.0}}
\begin{document}

\title{\setstretch{1.2}\Large\bfseries Market Efficiency in a Changing Asset Population:\\[2pt]
Mobility, Entry, and Survival over Cryptocurrency's Full History}
\author{
\setstretch{1.2}
\begin{tabular}[t]{c}
Antonis Ballis\thanks{Corresponding author. Department of Accounting and Finance, Aston Business School, Aston University, Aston Triangle, Birmingham B4 7ET, United Kingdom. E-mail: \texttt{a.ballis@aston.ac.uk}.}\\
{\normalsize Aston Business School, Aston University}
\end{tabular}
\and
\begin{tabular}[t]{c}
Christos Ioannidis\thanks{Department of Accounting and Finance, Aston Business School, Aston University, Aston Triangle, Birmingham B4 7ET, United Kingdom. E-mail: \texttt{c.ioannidis@aston.ac.uk}.}\\
{\normalsize Aston Business School, Aston University}
\end{tabular}}
\date{October 2026}
\maketitle
\thispagestyle{empty}

\begin{abstract}
\noindent\setstretch{1.0}
Weak-form efficiency implies symmetry restrictions on the Markov chain governing movement across the cryptocurrency capitalisation distribution. We test them on the full listed population: 8,970 assets, 690 weekly cross-sections, 2013 to 2026, with entry, exit and survival observed rather than inferred. Symmetry is rejected in every era, but the asymmetry falls by seventy percent, and what converged is entry-driven compositional churn rather than coins' own valuations. The bottom decile earns 27.5\% per week on paper and nothing among assets that trade, on capacity of \$293,000. Compensation for neglect exists precisely where trading does not.
\end{abstract}

\medskip
{\setstretch{1.15}
\par}

\newpage

\section{Introduction}
\label{sec:intro}

A prominent literature documents a large size premium in cryptocurrency returns \citep{liu2021risks,liu2022common}. A size gradient in expected returns is not, by itself, evidence against market efficiency. In equilibrium, small cryptocurrencies may earn more because they are less recognised and more likely to fail \citep{merton1987}, and large ones may earn less because holders derive transactional benefits from them \citep{cong2021tokenomics}. Any test of efficiency in this cross-section is therefore a joint test with a model of expected returns \citep{fama1970}. This paper is organised around that problem. Rather than asking whether small coins earn more, we ask two narrower questions whose answers do not depend on the benchmark: whether a coin's position in the capitalisation distribution predicts the direction of its next move, and whether any return gradient survives in the part of the market an investor can actually trade.

We answer both over the market's full history. We make the first question precise by modelling movements across capitalisation deciles as a finite-state Markov chain augmented with structural entry and absorbing exit states, following the mobility framework of \citet{ballis2026}, and by translating weak-form efficiency into two testable symmetry restrictions on that chain. The first is \emph{conditional symmetry}: from any decile, upward and downward transitions must be equally likely, otherwise the current decile predicts the direction of future movement. The second is \emph{reversibility} (detailed balance): the stationary probability flow between any two deciles must balance, which is the appropriate unconditional restriction when the stationary distribution is not uniform and the chain includes entry and exit. Both restrictions have exact count-based tests---a Wald test on directional contrasts of the transition matrix and a Bowker-type test on transition counts---and both connect directly to portfolio economics: under symmetry, payoff antisymmetry and exit hazards that do not vary across deciles, no decile-based trading rule can generate positive expected returns, a result we formalise below---and whose maintained payoff restriction we test, and reject, in Section~\ref{sec:payoffsym}.

We observe the population of capitalisation-reporting cryptocurrencies on CoinMarketCap at weekly frequency from April 2013 to July 2026: 690 weekly cross-sections, 1.21 million coin-week observations drawn from 2.95 million listed coin-weeks, and 8,970 distinct assets, keyed on stable numeric identifiers so that symbol reuse does not contaminate transitions. Assets are excluded when CoinMarketCap reports no circulating supply, and hence no capitalisation to rank, which accounts for 99.5\% of excluded rows; that fringe is large in identifiers but accounts for roughly one percent of trading volume, and Table~\ref{tab:coverage} characterises it in full. Within the capitalisation-reporting universe there is no survivorship selection \citep{brown1992}: births and deaths of assets are observed rather than inferred, and every asset is followed to its delisting. \citet{ballis2026} establish the market's era structure endogenously: least-squares break detection on a rolling mobility index locates structural breaks in November 2017 and July 2021, partitioning the market's history into a \emph{frontier}, an \emph{expansion}, and a \emph{consolidation} era. We adopt those breaks throughout.

Five results emerge. The first is statistical. Both symmetry restrictions are overwhelmingly rejected in every era, under asymptotic inference and under a permutation scheme robust to within-week cross-sectional dependence, so the cryptocurrency cross-section is not Markov mobility-efficient at any point in its history. But the economic magnitude declines steadily: the mean absolute gap between upward and downward transition probabilities falls from 0.040 in the frontier era to 0.028 and then 0.022, and a rolling version of the index peaks at 0.062 in late 2014 and ends the sample at 0.018, with a significant negative log-linear trend.

The second result concerns payoffs: the asymmetry maps into a large anomaly that was never a portfolio strategy. Among surviving coins an equal-weighted bottom-decile portfolio earns market-adjusted returns of 27.5\% per week in the frontier era and 2.8\% in the consolidation era. Successively more realistic frictions dismantle it. Treating delisted coins as total losses cuts the frontier premium by a quarter; screening on a token \$10,000 of daily volume eliminates it entirely ($t=0.19$), because the premium lived in coins that could not be traded---the investable bottom decile held on average 2.0 coins in only 35 tradable weeks of that era; and weekly turnover of 17--47\% erodes, without eliminating, what survives the screens. In the later eras the screened premium does survive, but only under a construction no investor faces; re-forming deciles inside the tradable universe---the cross-section an investor actually confronts---leaves a consolidation-era gradient of $0.01\%$ per week ($t=0.04$), and the portfolio that carries the screened premium has a median equal-weighted capacity of \$293,000. Section~\ref{sec:payoff} reports the details, including capacity estimates, skip-week timing, and factor adjustment, all of which sharpen rather than rescue the conclusion.

A separate line of evidence points the same way. The premium is not a property of the typical bottom-decile coin at all. Its cross-sectional \emph{median} is $+7.2\%$ per week in the frontier era but $-2.6\%$ in the expansion era and $-0.1\%$ in the consolidation era; the 10--90 trimmed means are $+15.3\%$, $-0.04\%$ and $-0.09\%$. Outside the frontier era the entire measured premium is generated by the right tail. Two measurement features produce that tail: the median bottom-decile coin is quoted at a fraction of a cent, so a one-tick move is a large percentage return, and 47\% of consolidation-era bottom-decile coin-weeks report zero trading volume, so their prices are stale and their returns arrive as accumulated jumps. Imposing a one-cent minimum price cuts the frontier premium from 27.6\% to 9.4\% per week.

The third result is a decomposition separating two sources of rank asymmetry. Measured against decile boundaries frozen at the start of each week---so that only a coin's own capitalisation movement matters---the asymmetry is small and essentially flat across eras (0.025--0.026). The component that converged is compositional: the churn induced by a very high entry rate (3.9\% of the universe per week in the frontier era, falling to 0.8\% in the consolidation era) that mechanically displaced incumbents through the relative ranking. Genuine own-valuation asymmetry was modest all along; what matured is the market's composition.

The fourth result concerns the structure that convergence leaves behind. Section~\ref{sec:framework} develops a parsimonious attention-and-survival framework in which information acquisition is costly, investor recognition rises with capitalisation \citep{merton1987}, and economic survival is itself uncertain. It requires no departure from rationality: limited attention is a rational response to processing constraints. The framework delivers three cross-sectional comparative-static restrictions---on rank persistence above a threshold decile $k^{*}$, on exit hazards, and on conditional survivor returns---together with a time-series implication about market development that is evaluated separately and does not enter the joint test. Formal many-inequality tests \citep{cck2014} fail to reject the three cross-sectional restrictions, individually or jointly, in every era.

The fifth result qualifies the fourth, and is the paper's most consequential single finding. Re-running the same tests on universes screened by trading volume separates the framework's predictions sharply. The exit-hazard ordering survives every screen and every decile construction, and the persistence ordering survives every screen once deciles are re-formed within the screened universe: larger assets remain more positionally stable and less likely to disappear whether or not one restricts attention to assets that trade. The conditional return ordering does not. In the consolidation era it is rejected at a \$10,000 volume screen ($p=0.010$) and at \$1{,}000{,}000 ($p=0.001$), and sits marginally above the threshold at \$100,000 ($p=0.053$); on tradable assets the size gradient in returns is absent. Compensation for neglect is therefore a property of the segment of the market that cannot be traded---a far narrower claim than the unrestricted prediction. It also dissolves what would otherwise be a contradiction between the anomaly result and the framework: the premium exists where trading does not, and vanishes where trading does.

We are correspondingly cautious about two further pieces of evidence. A time-series regression of asymmetry on market breadth does not survive conservative inference: breadth is nearly a deterministic trend over this sample, and in non-overlapping annual data it changes sign and loses significance while the entry rate dominates. And a difference-in-differences-style comparison around the spot-ETF approvals finds no event-time discontinuity, but the design is demonstrably underpowered, so we report it descriptively and rest nothing on it.

The paper contributes to three literatures. To the cryptocurrency asset-pricing literature, it adds a result that the existing evidence does not deliver. \citet{liu2022common} show that the size premium is concentrated among illiquid, costly-to-arbitrage coins but survives as a priced factor in a universe screened at \$1 million of capitalisation; our expansion era, which overlaps their sample, is consistent with that finding, and so is the growing cross-sectional factor evidence in this market \citep{fieberg2025}. Following the full population through 2026, with entry and exit observed, we find that in the consolidation era the premium is absent among assets that trade, that it is carried by the right tail rather than by the typical small coin, and that the size ordering of conditional returns implied by equilibrium compensation is rejected on tradable universes at the \$10,000 and \$1,000,000 volume screens. The premium is therefore not an illiquidity premium earned by investors in small coins; it is a property of a segment of the market in which no one can invest. The pattern is in the spirit of equity-market findings that anomaly returns concentrate in high-cost, hard-to-trade stocks \citep{novymarx2016,chordia2014}, attenuate after publication \citep{mclean2016}, and are intertwined with priced illiquidity \citep{amihud2002}. To the market-efficiency literature, it contributes a transition-based formulation of weak-form efficiency that tests the cross-sectional dimension of predictability---who moves where in the size distribution---rather than the time-series dimension studied by the runs-test tradition \citep{fama1970} and, in cryptocurrency markets, by evidence of return predictability, market segmentation, and price distortions \citep{liu2021risks,makarov2020,griffin2020}. To the mobility literature in the tradition of \citet{anderson1957}, it shows that mobility tools carry economic content beyond description: symmetry of the transition matrix is a sufficient condition for zero expected returns on decile-conditioned rules under payoff antisymmetry and decile-invariant exit hazards, and reversibility is its stationary counterpart. Because we reject payoff antisymmetry in Section~\ref{sec:payoffsym}, we keep the two objects separate throughout---directional mobility symmetry is a statement about transition probabilities, weak-form efficiency a statement about returns---and evaluate transition predictability and realised portfolio returns in turn. Their interaction, not either alone, determines whether statistical mobility asymmetry creates an economically exploitable opportunity.

\section{Mobility symmetry and weak-form efficiency}
\label{sec:theory}

\subsection{The capitalisation chain}
\label{sec:chain}

Let each coin $c$ alive in week $t$ occupy a state $S_{c,t}\in\{1,\dots,K\}$, $K=10$, given by its decile in the cross-sectional distribution of market capitalisation (decile 1 the smallest, decile 10 the largest). The state space is augmented with a structural entry state $N$ (newly listed coins enter from $N$) and an absorbing exit state $X$ (delisted coins move to $X$ and remain there). Assume $S_{c,t}$ follows a first-order, within-era time-homogeneous Markov chain with transition matrix $P=[p_{ij}]$; both assumptions are tested and discussed at length in \citet{ballis2026}, where deviations are found to be statistically detectable in samples of this size but economically modest and shrinking over time. Interior transition probabilities are understood conditional on survival, $p_{ij} = \Pr(S_{t+1}=j \mid S_t=i,\, S_{t+1}\neq X)$, with exit treated separately through the exit hazard $h_i = \Pr(S_{t+1}=X \mid S_t=i)$.

\subsection{Efficiency restrictions}

Associate with a transition from decile $i$ to decile $j$ a payoff $\mu_{ij}$, the expected return of a coin making that transition. Weak-form efficiency in the sense of \citet{fama1970} requires that the information in the current state does not predict returns. Because a coin may also leave the market entirely, the conditional expectation has two parts:
\begin{equation}
E\left(r_{t+1}\mid S_t=i\right)=\left(1-h_i\right)\sum_{j\le K} p_{ij}\,\mu_{ij} \;+\; h_i\,\mu_{iX} = 0 \quad \text{for every decile } i,
\label{eq:emh}
\end{equation}
where $h_i$ is the exit hazard of Section~\ref{sec:chain} and $\mu_{iX}$ is the return realised on delisting.

Equation~\eqref{eq:emh} makes two things immediately clear, and both discipline what follows. First, it is an unrestrictive statement until structure is placed on payoffs. Second, and less obviously, the interior term and the exit term must offset each other decile by decile. Setting $\mu_{iX}=-1$, as the harshest treatment of delisting in Section~\ref{sec:payoff} does, the interior expectation must equal $h_i/(1-h_i)$ rather than zero, and that quantity varies across $i$ whenever the exit hazard does. Symmetry of the interior chain therefore delivers weak-form efficiency only if the exit hazard is constant across deciles. It is not: Section~\ref{sec:persistence} measures hazards falling monotonically from 2.0\% per week in the bottom decile to 0.2\% in the top, and Prediction~2 of Section~\ref{sec:framework} asserts that ordering as an equilibrium feature rather than an accident. We therefore treat interior symmetry and return efficiency as two distinct objects throughout, and quantify the exit wedge directly in Section~\ref{sec:payoff}.

Equation~\eqref{eq:emh} sets expected returns to zero for exposition. The appropriate benchmark allows them to differ across deciles for equilibrium reasons: compensation for neglect and survival risk, developed in Section~\ref{sec:framework}, and a convenience yield on widely used coins, under which users accept lower financial returns in exchange for transactional benefits \citep{cong2021tokenomics}. Both channels predict that expected returns fall with size in an efficient market. We therefore never interpret a size premium as a rejection of efficiency. A rejection of~\eqref{eq:sym} below establishes predictability in rank dynamics; whether that predictability reflects mispricing, compensation, or measurement is decided by what happens at the investable margin (Section~\ref{sec:payoff}) and by the equilibrium restrictions tested in Section~\ref{sec:cck}.

Setting the exit term aside for a moment, under \emph{payoff antisymmetry}---the return to moving up $k$ deciles is minus the return to moving down $k$ deciles, $\mu_{i,i+k}=-\mu_{i,i-k}$, with $\mu_{ii}=0$, a maintained assumption tested directly in Section~\ref{sec:payoffsym}---the interior condition reduces to a restriction on transition probabilities alone:
\begin{equation}
H_0^{\mathrm{sym}}:\quad p_{i,i+k}=p_{i,i-k} \quad \text{for all interior } i \text{ and all feasible } k.
\label{eq:sym}
\end{equation}
Restriction~\eqref{eq:sym} states that from any decile, moves of equal size up and down are equally likely. We refer to it, and to its stationary counterpart below, as \emph{Markov mobility efficiency}: a property of the transition matrix alone, and a strictly narrower object than weak-form efficiency in returns. If~\eqref{eq:sym} holds, payoffs are antisymmetric, and exit hazards are constant across deciles, then for any portfolio formed on decile position expected one-step returns are zero, and no reweighting across deciles can produce a positive expected return. Conversely, a rejection of~\eqref{eq:sym} identifies deciles from which the direction of future movement is predictable; whether that direction is worth anything is a separate question, answered in Section~\ref{sec:payoff}.

The status of the payoff restriction requires care, because the paper's own evidence bears on it. Antisymmetry is not merely rejected in Section~\ref{sec:payoffsym}; it is close to unattainable in this cross-section, and would be so even under a null of no predictability whatsoever. Simple returns are bounded below at $-100\%$ and unbounded above, so a coin that ascends $k$ deciles can gain far more than a coin that descends $k$ deciles can lose. To gauge the size of that wedge we simulate a cross-section of 2{,}000 assets whose returns are drawn \emph{exactly symmetrically in logs}---median simple return zero, no drift, no state dependence, and hence no predictability of any kind---rank them, and assign decile moves. The resulting $\mu^U_k+\mu^D_k$ is positive at every $k$ and rises with $k$: at a weekly log volatility of $0.30$ it averages $+4.7\%$, $+8.5\%$ and $+30.3\%$ for $k=1,2,3$, and is positive in 99--100\% of simulated weeks; at $0.60$, plausible for micro-capitalisation crypto assets, the figures are $+20.2\%$, $+35.0\%$ and $+132.3\%$. This is the same sign and the same increasing-in-$k$ shape that Table~\ref{tab:paysym} reports in the data.

Two consequences follow. First, Section~\ref{sec:payoffsym} should be read as measuring the size of a wedge that the return distribution guarantees, not as adjudicating a substantive economic hypothesis. Second, the mapping from~\eqref{eq:sym} to weak-form efficiency, which requires that wedge to be zero, cannot be maintained. Accordingly we do not rest the efficiency claim on the symmetry tests. Sections~\ref{sec:symmetry} and~\ref{sec:payoff} evaluate transition predictability and realised portfolio returns separately: the first asks whether current position predicts the direction of movement, the second asks what that direction is worth once exit, illiquidity and costs are imposed. It is their conjunction, and specifically the direct estimate of $E(r_{t+1}\mid S_t=i)$ in Table~\ref{tab:payoff}, that carries the economic conclusion.

The unconditional counterpart of~\eqref{eq:sym} operates through the stationary distribution $\pi$ solving $\pi = \pi P$. Because the decile chain has a non-uniform stationary distribution and is embedded in a system with entry and exit, the natural stationary null is not symmetry of $P$ itself but \emph{reversibility} (detailed balance):
\begin{equation}
H_0^{\mathrm{rev}}:\quad \pi_i\, p_{ij} = \pi_j\, p_{ji} \quad \text{for all } i\neq j,
\label{eq:rev}
\end{equation}
which states that the stationary probability flow from $i$ to $j$ equals the reverse flow. Under~\eqref{eq:rev} the long-run distribution assigns no directional drift to any coin's traverse of the size distribution: paths up and their reversals are equally probable. Detailed balance equalises stationary probability \emph{flows}; combined with payoff antisymmetry and decile-invariant exit hazards it implies that the unconditional expected payoff of any decile-based rule is zero. Without that payoff restriction it does not, because forward and reverse transitions may carry payoffs of unequal magnitude---a possibility that Section~\ref{sec:payoffsym} shows is empirically the relevant one. Reversibility has an immediate empirical translation: since the expected count of $i\!\to\!j$ transitions in a stationary sample of $T$ transitions is $T\pi_i p_{ij}$, detailed balance holds if and only if expected transition counts are symmetric, $E[n_{ij}]=E[n_{ji}]$.

\subsection{Test statistics}

Let $n_{ij}$ denote observed transition counts over a given era, $n_i=\sum_j n_{ij}$ row totals, and $\hat p_{ij}=n_{ij}/n_i$ maximum-likelihood estimates \citep{anderson1957}. Reversibility is tested with the Bowker-type statistic
\begin{equation}
B \;=\; \sum_{i<j}\frac{\left(n_{ij}-n_{ji}\right)^2}{n_{ij}+n_{ji}} \;\;\sim\;\; \chi^2_{45},
\label{eq:bowker}
\end{equation}
with 45 degrees of freedom for the $10\times 10$ interior chain \citep{bowker1948}. For conditional symmetry, stack the aggregate directional contrast for each interior decile. Define, conditional on survival, the total upward and downward probabilities $p_i^{U}=\sum_{j>i}\hat p_{ij}$ and $p_i^{D}=\sum_{j<i}\hat p_{ij}$. Because each row of the transition matrix is estimated from an independent multinomial sample, the contrast $d_i=\hat p_i^{U}-\hat p_i^{D}$ has variance $\mathrm{Var}(d_i)=\left[p_i^{U}+p_i^{D}-(p_i^{U}-p_i^{D})^2\right]/n_i$, yielding a per-decile statistic $z_i = d_i/\sqrt{\widehat{\mathrm{Var}}(d_i)}$ and, by row independence, a joint Wald statistic $W=\sum_{i=2}^{9} z_i^2 \sim \chi^2_8$ over the eight interior deciles for which both directions are feasible. More generally, the full set of antisymmetric contrasts can be written $d = R\,\mathrm{vec}(\hat P)$ for a known selection matrix $R$ whose rows are $\mathrm{vec}(E_{ij}-E_{ji})^\top$, with $W=d^\top\!\left(R\,\Omega\, R^\top\right)^{-1} d$ and $\Omega$ the block-diagonal multinomial covariance of $\mathrm{vec}(\hat P)$; the aggregate directional version reported below is the economically focused special case.

Asymptotic inference treats transitions as independent, but coin moves within a week are cross-sectionally dependent---ranks are zero-sum. We therefore complement the $\chi^2$ $p$-values with a week-level permutation scheme: under the null, the \emph{direction} of each week's flow matrix is exchangeable, so we transpose each weekly interior count matrix with probability one-half, recompute the statistics, and repeat 999 times. With 999 replications the smallest attainable $p$-value is $1/1000$; permutation $p$-values are computed as $(1+\#\{\text{replications}\ge\text{observed}\})/1000$ and reported as $<$0.001 where no replication exceeds the observed statistic. Because entire weeks are flipped, all within-week dependence is preserved under resampling.

As a summary measure of economic magnitude---which, unlike the test statistics, does not scale mechanically with sample size---we report the mean absolute directional asymmetry
\begin{equation}
\mathcal{A} \;=\; \frac{1}{8}\sum_{i=2}^{9}\left|\,p_i^{U}-p_i^{D}\,\right|,
\label{eq:asym}
\end{equation}
computed era-by-era and on a rolling 52-week window.

\subsection{Mechanical considerations}
\label{sec:mechanical}

Three features of rank-based transitions require care. First, ranks are zero-sum: a coin ascending forces another to descend, so deviations from symmetry cannot arise from common shocks and reflect genuinely cross-sectional reallocation; the permutation inference above additionally guards against the dependence this induces. Second, the asset universe grows: entrants land mid-distribution and mechanically push incumbents down in relative rank, generating downward drift among survivors that reflects composition rather than mispricing. Section~\ref{sec:owncap} quantifies this channel directly by re-measuring transitions against decile boundaries frozen at the start of each week, so that only a coin's own capitalisation movement can generate a transition. Third, conditioning on survival truncates the left tail of the bottom deciles---coins that would have moved down exit instead---which biases the measured drift of low deciles upward. The payoff analysis of Section~\ref{sec:payoff} therefore treats delisting explicitly, assigning exiting coins a return of $-100\%$ in the harshest variant.

\subsection{What efficiency does not rule out}
\label{sec:risk}

Symmetry restricts first moments only. Two chains can both satisfy~\eqref{eq:sym} and yet differ radically in risk, because persistence---the staying probability $s_i=p_{ii}$---governs the temporal clustering of positions. Expected sojourn duration in decile $i$ is $1/(1-s_i)$, and multi-period risk measures are convex in persistence. The mechanism is elementary. If the per-period payoff attached to a state is persistent with first-order autocorrelation $s$, the variance of the $h$-period cumulative return grows faster than $h$: for large $h$ it scales with $h\,\sigma^2(1+s)/(1-s)$, the familiar long-run-variance inflation factor. A state with $s=0.95$ therefore carries roughly thirty-nine times the long-run variance of an otherwise identical state with $s=0$, at the same one-period volatility. Persistence converts symmetric one-period shocks into asymmetric multi-period costs, because long adverse sojourns dominate the tail. We do not attempt a welfare calculation; the point is only that first-moment symmetry places no restriction on this second-moment structure. The empirical counterpart in Section~\ref{sec:persistence} shows that what survives in the maturing cryptocurrency market is exactly this second-moment structure---persistence and tail risk that differ systematically across deciles and grow with horizon---rather than exploitable drift.

\section{Data}
\label{sec:data}

The dataset is the population of cryptocurrencies listed with reported market capitalisation on CoinMarketCap, sampled at every weekly (Sunday) snapshot from 28 April 2013---the first snapshot ever published---through 12 July 2026: 690 cross-sections and 2,945,997 coin-week rows, of which 1,213,568 enter the decile panel: 1,210,166 with directly reported positive capitalisation and price, plus 3,402 bridged rows carrying interpolated capitalisation. Coins absent from a single weekly snapshot but present in the adjacent weeks---0.3\% of coin-weeks, including a corrupted June 2023 snapshot from which roughly 500 coins temporarily disappear---are bridged with interpolated capitalisation; bridged observations enter the transition analysis but are excluded from all return calculations. Coins are keyed on CoinMarketCap's stable numeric identifier rather than ticker symbols, which are reused extensively after 2020. Each week, all coins with positive capitalisation are ranked and assigned to deciles; a coin present in week $t$ but absent in week $t+1$ is recorded as an exit, and a coin present in $t+1$ but not $t$ as an entry.

The gap between 2,945,997 listed coin-weeks and the 1,213,568 that enter the decile panel is large enough to require its own accounting, and Table~\ref{tab:coverage} provides it. Essentially all of the difference---99.5\% of excluded rows---is CoinMarketCap reporting a capitalisation of exactly zero, which occurs when circulating supply is unknown or unverified. Such rows cannot be ranked in a capitalisation distribution, so exclusion is forced rather than chosen. Three features of the excluded set matter for interpretation.

First, the excluded observations are not simply dead assets. Ninety-one percent carry a positive quoted price, 74\% report positive 24-hour volume, and 36\% report volume above \$10,000---against 57\% among included rows. Their median 24-hour volume is \$910, against \$31,000 for included rows: much smaller, but not inert. Second, the exclusion rate is strongly time-varying, rising from 14\% of listed coin-weeks in the frontier era to 68\% in the consolidation era as listing volume outgrew supply verification. Third, and consequently, the panel covers 8,970 of the 33,797 distinct identifiers that appear in the snapshots; 24,827 never report a capitalisation and so never enter.

What limits the damage is that the excluded fringe is economically small on the dimension that matters for the paper's conclusions. Coins that never enter the panel account for 0.98\% of aggregate 24-hour trading volume over the full sample and 1.1\% after 2021. The excluded share of volume is 1.9\% over the full sample and never exceeds 2.2\% in any era. The decile panel is therefore best described not as the population of listed identifiers, but as the population of \emph{capitalisation-reporting} assets, which is the population the mobility question is posed over and which accounts for approximately 98\% of trading activity. We use that description throughout. Within it, and this is the property that matters for the payoff analysis, there is no survivorship selection: every asset that ever reports a capitalisation is observed in every subsequent week until it delists, and its delisting is recorded rather than inferred \citep{brown1992}. The caveat is that assets which never report supply are outside the frame entirely, and no result here speaks to them.

Table~\ref{tab:desc} summarises the three eras. The three eras differ sharply on every dimension: the average weekly cross-section grows from 441 coins in the frontier era to 2,885 in the consolidation era; average total capitalisation grows hundredfold; the weekly entry rate falls from 3.9\% of the universe to 0.8\% and the exit rate from 1.9\% to 0.6\%. The median listed coin remains tiny throughout (\$60,000 in the frontier era, \$2.6 million in the consolidation era)---a first indication that the typical member of this population is far below institutional investability thresholds. The share of listed coins reporting zero 24-hour volume rises from near nil to 14\% over the sample: a large and growing fringe of listed-but-dead assets that the investability analysis exploits.

\begin{table}[t]
\centering
\begin{threeparttable}
\caption{The three eras of the cryptocurrency market: descriptive statistics}
\label{tab:desc}
\scriptsize
\setlength{\tabcolsep}{6pt}
\begin{tabular}{lccccccc}
\toprule
Era & Weeks & Coins/wk & Cap (\$bn) & Med.\ cap (\$m) & Entry (\%/wk) & Exit (\%/wk) & Zero vol.\ (\%) \\
\midrule
Frontier & 239 & 441 & 22.6 & 0.06 & 3.93 & 1.85 & 1.2 \\
Expansion & 190 & 1,870 & 489.6 & 1.29 & 0.98 & 0.47 & 11.5 \\
Consolidation & 261 & 2,885 & 2,250.1 & 2.62 & 0.75 & 0.56 & 14.1 \\
\bottomrule
\end{tabular}
\begin{tablenotes}
\footnotesize
\item Notes: Coins/week is the average number of listed coins with positive reported capitalisation; total capitalisation is the within-era average of weekly totals; median cap is the median coin-week capitalisation. Entry (exit) is the average weekly share of the universe newly listed (delisted). Zero volume is the share of coin-weeks reporting zero 24-hour volume; volume is not tracked in the 2013 snapshots.
\end{tablenotes}
\end{threeparttable}
\end{table}

A companion volume panel, collected from the same snapshots, records 24-hour dollar trading volume for every coin-week; volume is unavailable in the 2013 snapshots. Weekly returns are computed from prices for coins surviving to the next snapshot and winsorised cross-sectionally at the 1st and 99th percentiles each week to contain data errors in micro-capitalisation coins. Era boundaries throughout are the endogenous structural breaks of \citet{ballis2026}: 19 November 2017 and 11 July 2021, defining the frontier (2013:04--2017:11), expansion (2017:11--2021:07), and consolidation (2021:07--2026:07) eras. Macro-financial covariates (VIX, federal funds rate) are from FRED; Bitcoin dominance, stablecoin share, and market breadth are computed from the panel itself.

\begin{table}[t]
\centering
\begin{threeparttable}
\caption{Sample coverage: listed coin-weeks, the decile panel, and what is excluded}
\label{tab:coverage}
\scriptsize
\setlength{\tabcolsep}{5pt}
\begin{tabular}{lccccc}
\toprule
& Listed & In decile & Excluded & Excluded share of & Excluded with \\
Era & coin-weeks & panel & (\%) & 24h volume (\%) & vol.\ $>$ \$10k (\%) \\
\midrule
Frontier (2013:04--2017:11)      & 120,448   & 103,252   & 14.3 & 1.0 & 11.7 \\
Expansion (2017:11--2021:07)     & 498,988   & 354,677   & 28.9 & 1.3 & 47.5 \\
Consolidation (2021:07--2026:07) & 2,326,561 & 752,237   & 67.7 & 2.2 & 34.8 \\
\midrule
Full sample                      & 2,945,997 & 1,210,166 & 58.9 & 1.9 & 35.6 \\
\bottomrule
\end{tabular}
\begin{tablenotes}
\footnotesize
\item Notes: A coin-week enters the decile panel if CoinMarketCap reports both a strictly positive market capitalisation and a strictly positive price. Of excluded rows, 99.5\% report a capitalisation of exactly zero, which occurs when circulating supply is unknown or unverified; the remaining 0.5\% lack a valid price. Excluded rows are nonetheless mostly quoted assets: 90.7\% carry a positive price and 74.4\% positive 24-hour volume, with a median 24-hour volume of \$910 against \$31,323 for included rows. Column 5 reports the share of excluded rows whose 24-hour volume exceeds \$10,000; the corresponding figure for included rows is 57.3\%. Across all snapshots 33,797 distinct identifiers appear, of which 8,970 report a capitalisation at least once and so enter the panel; the 24,827 that never do account for 0.98\% of aggregate 24-hour volume over the full sample and 1.1\% after July 2021. The panel row count of 1,210,166 in this table counts strictly positive capitalisation and price; the 1,213,568 rows used in the transition analysis additionally include 3,402 bridged observations, which carry interpolated capitalisation and are excluded from all return calculations.
\end{tablenotes}
\end{threeparttable}
\end{table}

\section{Symmetry tests and convergence}
\label{sec:symmetry}

\begin{figure}[t]
\centering
\includegraphics[width=\textwidth]{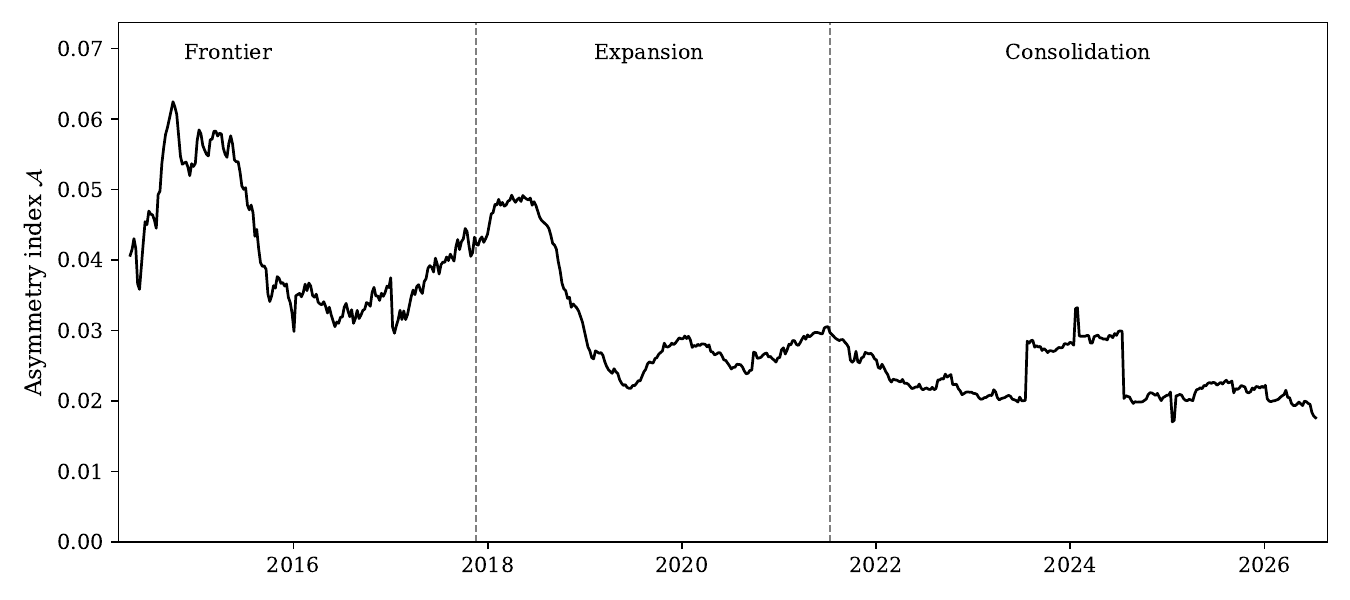}
\caption{Rolling directional asymmetry. The figure plots the 52-week rolling mean absolute gap between upward and downward transition probabilities across interior deciles ($\mathcal{A}$ of equation~\eqref{eq:asym}). Dashed vertical lines mark the era breaks of 19 November 2017 and 11 July 2021. Weekly data, 28 April 2013 to 12 July 2026, 690 cross-sections; source CoinMarketCap snapshots.}
\label{fig:asym}

\end{figure}

Table~\ref{tab:sym} reports the reversibility statistic~\eqref{eq:bowker} and the joint directional Wald statistic, era by era, with both asymptotic and permutation $p$-values, together with the magnitude index $\mathcal{A}$ of equation~\eqref{eq:asym}. Both nulls are rejected at any conventional level in every era, and the permutation $p$-values---which preserve all within-week cross-sectional dependence---confirm that the rejections are not an artefact of the independence assumption. With row counts of ten thousand transitions per decile in the frontier era rising to seventy-five thousand in the consolidation era, and around one hundred and twenty thousand per decile in the pooled sample, statistical rejection is unsurprising; the economically meaningful object is the magnitude. Here the pattern is unambiguous: $\mathcal{A}$ falls monotonically from 0.040 in the frontier era to 0.028 in the expansion era and 0.022 in the consolidation era. Figure~\ref{fig:asym} plots the rolling 52-week version of $\mathcal{A}$: it peaks at 0.062 in October 2014 and ends the sample at 0.018---a seventy percent decline from peak---and a log-linear trend regression on the rolling index yields a slope of $-0.14\%$ per week (Newey--West $t=-9.0$). Taken literally this implies a half-life of roughly ten years, but the extrapolation should not be taken literally: the index is bounded below at zero, so a log-linear form must eventually fail, and the regression is run on overlapping 52-week windows, so its standard error understates uncertainty. We use the trend as evidence that convergence is continuous rather than a step function, not as a forecast; the decline is a nearly continuous path rather than an artefact of the era partition.

The one visible interruption is a hump in the rolling index between late 2023 and mid-2024, when it temporarily rises from about 0.021 to 0.033 before resuming its decline. Inspection of the underlying weeks traces the episode to identifiable recomposition events on the listing platform rather than to gradual market dynamics: a batch delisting of 480 coins in the week of 16 July 2023---distinct from the corrupted June 2023 snapshot noted in Section~\ref{sec:data}, since the universe falls from 2,582 to 2,106 assets and does not recover, whereas the June episode reverses within a week and is bridged out---and batch listings of 403 and 273 coins in January and February 2024, each of which mechanically shifts the decile assignment of incumbents in a single week (weekly asymmetry readings of 0.37, 0.41 and 0.64 in the three affected weeks, against a consolidation-era median of 0.024). The 52-week window then carries these isolated weeks until they roll out of the sample. The episode is thus the composition channel of Section~\ref{sec:mechanical} in concentrated form; consistent with this reading, the asymmetry index co-moves with the entry rate---a full-sample correlation of 0.49 in unsmoothed weekly data, and 0.97 between the two 52-week rolling series, though the latter is inflated by the overlapping windows and should not be read as independent evidence---anticipating the decomposition of Section~\ref{sec:owncap} and the time-series evidence of Section~\ref{sec:drivers}.

Table~\ref{tab:zstats} reports the per-decile statistics $z_i$. Two features stand out. First, in every era, survivors in middle and upper deciles (3--9) drift downward---the composition channel of Section~\ref{sec:mechanical} at work as an expanding universe pushes incumbents down the relative ranking---while decile~2 drifts upward, reflecting in part survivorship truncation at the bottom. Second, the underlying probability gaps, summarised by $\mathcal{A}$, shrink across eras in five of the eight interior deciles and in the average; deciles 3, 4 and 6 fall between the frontier and expansion eras and then edge back, by roughly 0.001 in each case. Section~\ref{sec:owncap} shows that the shrinking component is precisely the compositional one.

\begin{table}[t]
\centering
\begin{threeparttable}
\caption{Reversibility and directional symmetry tests by era}
\label{tab:sym}
\scriptsize
\setlength{\tabcolsep}{6pt}
\begin{tabular}{lccccccc}
\toprule
& \multicolumn{3}{c}{Reversibility (Bowker)} & \multicolumn{3}{c}{Directional symmetry} & \\
\cmidrule(lr){2-4}\cmidrule(lr){5-7}
Era & $B$ & $p$ (asy.) & $p$ (perm.) & $W$ & $p$ (asy.) & $p$ (perm.) & $\mathcal{A}$ \\
\midrule
Frontier (2013:04--2017:11) & 213.5 & $<$0.001 & $<$0.001 & 577.0 & $<$0.001 & 0.001 & 0.0399 \\
Expansion (2017:11--2021:07) & 695.4 & $<$0.001 & $<$0.001 & 1,348.4 & $<$0.001 & 0.001 & 0.0282 \\
Consolidation (2021:07--2026:07) & 1,652.0 & $<$0.001 & $<$0.001 & 2,548.3 & $<$0.001 & 0.001 & 0.0218 \\
Full sample & 2,412.6 & $<$0.001 & $<$0.001 & 4,396.4 & $<$0.001 & 0.001 & 0.0252 \\
\bottomrule
\end{tabular}
\begin{tablenotes}
\footnotesize
\item Notes: $B$ is the Bowker statistic of equation~\eqref{eq:bowker} testing detailed balance on the interior $10\times10$ chain ($\chi^2_{45}$ under the null). $W$ is the joint Wald statistic $\sum_i z_i^2$ over interior deciles 2--9 testing $p_i^U=p_i^D$ conditional on survival ($\chi^2_{8}$). Permutation $p$-values transpose each weekly interior count matrix with probability one-half (999 replications), preserving within-week cross-sectional dependence. $\mathcal{A}$ is the mean absolute directional asymmetry of equation~\eqref{eq:asym}. Transition counts from 689 weekly transitions, April 2013--July 2026; era boundaries 19 November 2017 and 11 July 2021 \citep{ballis2026}.
\end{tablenotes}
\end{threeparttable}
\end{table}

\begin{table}[t]
\centering
\begin{threeparttable}
\caption{Per-decile directional asymmetry: $z$-statistics for $H_0\!: p_i^U=p_i^D$}
\label{tab:zstats}
\small
\begin{tabular}{lcccc}
\toprule
Decile & Frontier & Expansion & Consolidation & Full sample \\
\midrule
2 & 4.51$^{***}$ & 8.66$^{***}$ & 5.67$^{***}$ & 10.94$^{***}$ \\
3 & $-$2.36$^{**}$ & $-$3.56$^{***}$ & $-$7.27$^{***}$ & $-$8.13$^{***}$ \\
4 & $-$5.50$^{***}$ & $-$8.43$^{***}$ & $-$15.66$^{***}$ & $-$18.15$^{***}$ \\
5 & $-$6.79$^{***}$ & $-$11.42$^{***}$ & $-$17.21$^{***}$ & $-$21.55$^{***}$ \\
6 & $-$7.91$^{***}$ & $-$10.99$^{***}$ & $-$20.82$^{***}$ & $-$24.23$^{***}$ \\
7 & $-$9.58$^{***}$ & $-$15.04$^{***}$ & $-$20.16$^{***}$ & $-$26.84$^{***}$ \\
8 & $-$11.38$^{***}$ & $-$16.38$^{***}$ & $-$19.11$^{***}$ & $-$27.59$^{***}$ \\
9 & $-$13.81$^{***}$ & $-$21.07$^{***}$ & $-$26.78$^{***}$ & $-$36.72$^{***}$ \\
\bottomrule
\end{tabular}
\begin{tablenotes}
\footnotesize
\item Notes: $z_i=(\hat p_i^U-\hat p_i^D)/\mathrm{se}$, conditional on survival; positive values indicate upward drift. Decile 1 (bottom) and 10 (top) are omitted because only one direction is feasible. $^{*}$, $^{**}$, $^{***}$ denote significance at the 10\%, 5\%, and 1\% levels.
\end{tablenotes}
\end{threeparttable}
\end{table}

\subsection{Own-capitalisation versus compositional asymmetry}
\label{sec:owncap}

To separate movements caused by a coin's own valuation from movements caused by the changing composition of the universe, we re-measure every transition against decile boundaries \emph{frozen} at the start of the week: coin $c$'s destination state is the decile its week-$t+1$ capitalisation would occupy in the week-$t$ boundaries. Under this convention, entry and exit of other coins cannot move $c$; only $c$'s own capitalisation can. The resulting own-capitalisation asymmetry is small and essentially flat across eras---0.025 in the frontier era, 0.026 in the expansion era, 0.026 in the consolidation era (Table~\ref{tab:variants})---while the baseline relative-rank asymmetry falls from 0.040 to 0.022. If anything the own-capitalisation component edges upward while the relative-rank component halves---the strongest available form of the decomposition's conclusion. The convergence documented above is therefore driven by the compositional component: the torrential entry margin of the early market (Table~\ref{tab:desc}) mechanically churned incumbents through the ranking, and as entry decelerated the churn subsided. Genuine own-valuation directional asymmetry was modest throughout the market's history---an observation that already foreshadows the payoff results: predictability of \emph{relative position} need not translate into large tradable returns once composition is held fixed.

\section{Payoffs: the paper premium and the investable margin}
\label{sec:payoff}

Statistical asymmetry matters economically only if it maps into returns. Each week we sort all coins with positive capitalisation into deciles, form equal-weighted portfolios, and measure market-adjusted returns (portfolio return minus the equal-weighted market return) over the following week. Inference uses Newey--West standard errors \citep{newey1987} with four lags on the weekly portfolio return series.

\subsection{The paper premium}

Panel~A of Table~\ref{tab:payoff} reports the baseline among surviving coins. The pattern is a size effect of extraordinary magnitude: the bottom decile earns market-adjusted returns of 27.5\% per week in the frontier era, 8.0\% in the expansion era, and 2.8\% in the consolidation era, with the top-minus-bottom spread significantly negative in every era. Returns decline monotonically across deciles in the expansion era; in the consolidation era the gradient is monotone only through decile 6, after which it turns and rises modestly to $-0.13\%$ at decile 10. Taken at face value, this is an order of magnitude larger than the cryptocurrency size premium reported by \citet{liu2022common} on tradable samples, and it shrinks by an order of magnitude across eras---the payoff counterpart of the convergence in Figure~\ref{fig:asym}.

Two features of Panel~A should be noted immediately, because both anticipate the argument of this section. The first is that the frontier-era gradient is not monotone: deciles 2, 3 and 4 return 8.52\%, 0.12\% and 10.34\% respectively, and the decile-4 estimate carries a $t$-statistic of 0.91. Only the bottom decile is separated from the rest with any precision in that era, and the ordering among the middle deciles is not estimable. The second is that these are cross-sectional means; Section~\ref{sec:central} reports the same portfolios by median and trimmed mean.

\subsection{Delisting}

Panel~B treats every exit as a total loss: a coin absent from the following week's snapshot is assigned a return of $-100\%$. This is deliberately harsh (some exits are migrations or relistings), yet it bounds the survivorship channel of Section~\ref{sec:mechanical}. The frontier-era bottom-decile premium falls from 27.5\% to 20.4\% per week and its $t$-statistic from 3.7 to 2.8; the consolidation-era premium falls from 2.8\% to 1.6\%. Exit risk is concentrated exactly where the premium is: the full-sample weekly exit hazard is 2.0\% in the bottom decile versus 0.2\% in the top (Table~\ref{tab:persistence}).

\subsection{The investable margin}
\label{sec:invmargin}

Panel~C imposes a minimal investability screen: a coin must report more than \$10,000 of daily volume to be held. In the frontier era the consequence is immediate. The premium does not merely shrink---it collapses to statistical zero ($2.7\%$ per week, $t=0.19$), because the investable bottom decile contained on average 2.0 coins and there were only 35 weeks in the entire era in which such a portfolio could be formed at all. The premium of Panel~A lived in coins that could not be bought.

In the two later eras Panel~C tells a different story. The screened bottom decile earns 6.0\% per week in the expansion era ($t=2.63$) and 1.97\% in the consolidation era ($t=5.45$), the latter on a portfolio averaging 44.8 names and formable in 260 of 261 weeks. Taken at face value that is a large, diversified, persistently available anomaly. Two features of the estimate account for it: the decile construction, and the capacity it implies. Together they dispose of it.

\textbf{The decile construction.} Panel~C sorts the whole universe into deciles and then applies the screen. Its bottom decile is therefore not the small end of the tradable market; it is the smallest tenth of all listed assets that nonetheless trade \$10,000 a day---coins that are unusually liquid for their size, a group defined by the interaction of the sort and the screen rather than by either alone. The question an investor asks is different: among the assets I can trade, do the smaller ones earn more? Table~\ref{tab:tradable} answers it by re-forming deciles \emph{within} the tradable universe each week. The answer in the consolidation era is no. The spread between the largest and smallest tradable deciles is $0.01\%$ per week ($t=0.04$), and the cross-section is not even monotone: it is shallowly U-shaped, with the bottom decile at $+0.45\%$ ($t=2.60$), the middle deciles slightly negative, and the top decile at $+0.46\%$ ($t=3.34$). The expansion era retains a genuine tradable gradient ($-2.32\%$, $t=-2.14$); the frontier era's point estimate has the wrong sign and is insignificant ($+3.41\%$, $t=1.35$) on a bottom decile of five names. The size premium among tradable cryptocurrencies is thus an expansion-era phenomenon that had disappeared by the consolidation era, not a standing feature of the market.

\textbf{Capacity.} Even granting Panel~C its own construction, Table~\ref{tab:capacity} shows what the strategy is worth. Weekly turnover of 17\% at two percent one-way costs leaves 1.28\% per week in the consolidation era---costs discipline the premium but do not eliminate it. What eliminates it is capacity. Holding participation to a tenth of the thinnest constituent's daily volume, the equal-weighted portfolio carries a median of \$293,000, and 1.28\% per week on \$293,000 is \$3,757. In the frontier era the same calculation yields \$62 per week. A net return of 1.28\% per week on \$293,000 is not an investment opportunity at institutional scale. The returns are not statistically fragile---in the consolidation era they are not---but the vehicle carrying them cannot be scaled. At the \$100,000 screen the net-of-cost return is already negative in the consolidation era, and at \$1 million it is $-3.7\%$ per week.

The two arguments are independent and they point the same way. Constructed as an investor would construct it, the consolidation-era tradable cross-section has no size gradient; constructed on full-universe deciles and held to the equal-weighted participation bound, it has one worth a few thousand dollars a week, rising to roughly \$36,000 under the loosest, value-weighted bound reported below. We use the phrase \emph{illiquidity mirage} for this conjunction, and we mean it in a specific sense: not that the statistical premium is absent, but that it is a property of assets selected for being small relative to their liquidity, and it disappears both when the cross-section is re-formed among tradable assets and when the resulting portfolio is asked to hold money.

Table~\ref{tab:robust} pushes the analysis along two further margins, timing and screen severity. A skip-week convention---form the portfolio in week $t$, measure returns over week $t+1$ to $t+2$---removes bid--ask bounce; the loose-screen premium survives it in the expansion (5.2\%, $t=3.3$) and consolidation (1.1\%, $t=3.2$) eras, so the later-era predictability in Panel~C is not purely a microstructure artefact. Tightening the screen to \$100,000 leaves the same-week premium insignificant in both eras where it can be measured, though the skip-week consolidation cell remains significant (1.92\%, $t=3.27$); at \$1 million the estimate is negative in the consolidation era, the only era in which enough weeks survive to estimate it, on portfolios averaging 1.0--1.5 coins.

\begin{sidewaystable}[p]
\centering
\begin{threeparttable}
\caption{Market-adjusted weekly decile returns (\%)}
\label{tab:payoff}
\scriptsize
\setlength{\tabcolsep}{4pt}
\begin{tabular}{lccccccccccc}
\toprule
Era & D1 & D2 & D3 & D4 & D5 & D6 & D7 & D8 & D9 & D10 & D10$-$D1 \\
\midrule
\multicolumn{12}{l}{\emph{Panel A: surviving coins}}\\[4pt]
Frontier & 27.49 & 8.52 & 0.12 & 10.34 & $-$4.16 & $-$6.81 & $-$7.48 & $-$8.79 & $-$10.40 & $-$10.27 & $-$37.76$^{***}$ \\
 & (3.69) & (1.84) & (0.09) & (0.91) & ($-$1.92) & ($-$1.57) & ($-$1.79) & ($-$2.01) & ($-$1.96) & ($-$1.82) & ($-$2.91) \\[4pt]
Expansion & 8.05 & 3.85 & 1.15 & 0.03 & $-$1.01 & $-$1.57 & $-$2.19 & $-$2.65 & $-$2.66 & $-$2.81 & $-$10.86$^{***}$ \\
 & (9.60) & (5.72) & (3.61) & (0.12) & ($-$4.04) & ($-$7.63) & ($-$7.09) & ($-$5.70) & ($-$6.24) & ($-$4.86) & ($-$8.06) \\[4pt]
Consolidation & 2.78 & 0.86 & $-$0.10 & $-$0.54 & $-$0.70 & $-$0.80 & $-$0.62 & $-$0.39 & $-$0.31 & $-$0.13 & $-$2.91$^{***}$ \\
 & (12.05) & (5.45) & ($-$0.85) & ($-$6.32) & ($-$8.58) & ($-$10.32) & ($-$6.43) & ($-$3.51) & ($-$2.34) & ($-$0.88) & ($-$8.39) \\[4pt]
\midrule
\multicolumn{12}{l}{\emph{Panel B: delisting treated as $-100\%$}}\\[4pt]
Frontier & 20.41 & 6.98 & $-$0.88 & 10.44 & $-$3.62 & $-$5.78 & $-$6.06 & $-$7.26 & $-$8.53 & $-$8.32 & $-$28.73$^{**}$ \\
 & (2.84) & (1.50) & ($-$0.59) & (0.91) & ($-$1.71) & ($-$1.34) & ($-$1.45) & ($-$1.68) & ($-$1.62) & ($-$1.48) & ($-$2.27) \\[4pt]
Expansion & 6.91 & 3.40 & 1.06 & $-$0.04 & $-$0.82 & $-$1.43 & $-$1.96 & $-$2.37 & $-$2.30 & $-$2.42 & $-$9.33$^{***}$ \\
 & (8.06) & (5.18) & (3.23) & ($-$0.15) & ($-$3.16) & ($-$5.91) & ($-$5.88) & ($-$5.04) & ($-$5.45) & ($-$4.22) & ($-$6.90) \\[4pt]
Consolidation & 1.62 & 0.35 & $-$0.26 & $-$0.51 & $-$0.55 & $-$0.60 & $-$0.34 & $-$0.00 & 0.06 & 0.24 & $-$1.38$^{***}$ \\
 & (5.40) & (1.92) & ($-$1.80) & ($-$5.64) & ($-$6.27) & ($-$6.86) & ($-$3.15) & ($-$0.01) & (0.39) & (1.44) & ($-$3.13) \\[4pt]
\midrule
\multicolumn{12}{l}{\emph{Panel C: liquid coins only (24h volume $>$ \$10{,}000)}}\\[4pt]
Frontier & 2.71 & $-$10.84 & $-$3.14 & $-$3.81 & $-$14.77 & $-$2.35 & $-$5.57 & $-$1.60 & $-$0.56 & 0.56 & $-$3.89 \\
 & (0.19) & ($-$0.98) & ($-$0.40) & ($-$0.52) & ($-$3.47) & ($-$0.50) & ($-$1.96) & ($-$0.96) & ($-$0.45) & (1.37) & ($-$0.27) \\[4pt]
Expansion & 6.01 & 2.06 & 0.39 & 1.18 & 0.77 & 0.49 & $-$0.25 & $-$0.45 & $-$0.61 & $-$0.72 & $-$6.71$^{***}$ \\
 & (2.63) & (1.96) & (0.51) & (2.17) & (1.66) & (1.30) & ($-$1.31) & ($-$1.58) & ($-$2.82) & ($-$1.70) & ($-$2.63) \\[4pt]
Consolidation & 1.97 & 0.47 & $-$0.14 & $-$0.33 & $-$0.24 & $-$0.33 & $-$0.22 & 0.14 & 0.15 & 0.35 & $-$1.62$^{***}$ \\
 & (5.45) & (2.12) & ($-$0.71) & ($-$2.77) & ($-$2.17) & ($-$4.14) & ($-$2.64) & (1.59) & (1.39) & (2.80) & ($-$3.86) \\[4pt]
\bottomrule
\end{tabular}
\begin{tablenotes}
\footnotesize
\item Notes: Market-adjusted weekly returns of equal-weighted decile portfolios, in percent; Newey--West $t$-statistics (4 lags) in parentheses. Returns are winsorised cross-sectionally at the weekly 1st and 99th percentiles. Panel~C sorts the whole universe into deciles and then keeps the coin-weeks clearing a \$10,000 volume screen, so its D1 is the smallest tenth of all listed assets that happen also to trade. Table~\ref{tab:tradable} re-forms deciles \emph{within} the tradable universe instead, which is the portfolio an investor could actually assemble; the comparison between the two is the subject of Section~\ref{sec:invmargin}. $^{*}$, $^{**}$, $^{***}$: 10\%, 5\%, 1\%.
\end{tablenotes}
\end{threeparttable}
\end{sidewaystable}

\begin{table}[t]
\centering
\begin{threeparttable}
\caption{Decile returns among tradable assets}
\label{tab:tradable}
\scriptsize
\setlength{\tabcolsep}{3pt}
\begin{tabular}{lccccccccccc}
\toprule
Era & D1 & D2 & D3 & D4 & D5 & D6 & D7 & D8 & D9 & D10 & D10$-$D1 \\
\midrule
Frontier & $-$2.99 & 1.44 & 1.40 & $-$0.82 & $-$0.77 & 0.69 & $-$0.50 & 0.07 & 1.87 & 0.42 & 3.41 \\
 & ($-$1.42) & (1.02) & (1.11) & ($-$0.58) & ($-$0.80) & (0.68) & ($-$0.53) & (0.07) & (1.88) & (0.56) & (1.35) \\[4pt]
Expansion & 1.81 & 1.31 & 0.37 & $-$0.04 & $-$0.61 & $-$0.34 & $-$0.70 & $-$0.46 & $-$0.85 & $-$0.51 & $-$2.32$^{**}$ \\
 & (2.82) & (2.70) & (1.10) & ($-$0.22) & ($-$2.44) & ($-$1.11) & ($-$3.56) & ($-$1.85) & ($-$2.82) & ($-$1.00) & ($-$2.14) \\[4pt]
Consolidation & 0.45 & $-$0.38 & $-$0.21 & $-$0.26 & $-$0.39 & $-$0.09 & 0.13 & 0.18 & 0.12 & 0.46 & 0.01 \\
 & (2.60) & ($-$3.12) & ($-$1.89) & ($-$2.41) & ($-$4.69) & ($-$1.14) & (1.28) & (1.93) & (0.93) & (3.34) & (0.04) \\
\bottomrule
\end{tabular}
\begin{tablenotes}
\footnotesize
\item Notes: As Table~\ref{tab:payoff}, except that deciles are re-formed each week \emph{within} the universe of coins reporting more than \$10,000 of 24-hour volume, rather than formed on the whole universe and then screened. This is the cross-section an investor faces: D1 is the small end of the tradable market, not the smallest tenth of all listed assets that happen also to trade. Weeks containing fewer than 20 tradable assets are dropped, which is why this table covers 186 frontier weeks against the 35 in which a screened bottom decile of the Table~\ref{tab:payoff} Panel~C construction existed at all. (Table~\ref{tab:liquid} reports 185 frontier weeks at the same screen, one fewer, because a transition test additionally requires the following week to qualify.) The bottom decile averages 5, 85 and 181 names across the three eras. Newey--West $t$-statistics (4 lags) in parentheses; $^{*}$, $^{**}$, $^{***}$: 10\%, 5\%, 1\%.
\end{tablenotes}
\end{threeparttable}
\end{table}

\begin{table}[t]
\centering
\begin{threeparttable}
\caption{Bottom-decile returns net of costs, and portfolio capacity}
\label{tab:capacity}
\footnotesize
\setlength{\tabcolsep}{4pt}
\begin{tabular}{llcccccc}
\toprule
Screen & Era & Gross & ($t$) & Turnover & Net of & Capacity & Profit \\
 & & (\%/wk) & & (\%/wk) & 2\% costs (\%/wk) & (\$) & (\$/wk) \\
\midrule
$>$\$10k & Frontier & 2.71 & (0.19) & 46 & 0.87 & 7,100 & 62 \\
$>$\$10k & Expansion & 6.01 & (2.63) & 42 & 4.34 & 25,200 & 1,091 \\
$>$\$10k & Consolidation & 1.97 & (5.45) & 17 & 1.28 & 292,700 & 3,757 \\
\midrule
$>$\$100k & Expansion & 2.66 & (0.48) & 43 & 0.95 & 67,600 & 640 \\
$>$\$100k & Consolidation & 0.83 & (1.14) & 23 & $-$0.08 & 575,400 & $-$451 \\
\midrule
$>$\$1m & Consolidation & $-$2.71 & ($-$1.14) & 24 & $-$3.69 & 917,400 & $-$33,868 \\
\bottomrule
\end{tabular}
\begin{tablenotes}
\footnotesize
\item Notes: The bottom decile is Panel~C of Table~\ref{tab:payoff}---full-universe deciles, then the volume screen---which is the construction most favourable to the strategy. Gross is the market-adjusted weekly return with Newey--West $t$ (4 lags). Turnover is the weekly share of the portfolio replaced. Net of costs subtracts $2\tau c$ with $c=2\%$ one-way, charged on both legs. Capacity is the median across weeks of $0.1\,N\,\underline{v}/\tau$, where $N$ is the number of holdings, $\underline{v}$ the smallest constituent's daily dollar volume, and participation is capped at 10\% of that volume; it is the dollar amount the \emph{equal-weighted} portfolio can carry without exceeding the cap in its thinnest name. Profit is the net return applied to capacity. Cells with fewer than 20 usable weeks are omitted. A one-percent cost assumption raises every net-of-cost figure by $2\tau$ percentage points---0.3 to 0.9 points across the rows shown---and changes no sign except at the \$100k consolidation cell, which turns from $-0.08$ to $+0.38$.
\end{tablenotes}
\end{threeparttable}
\end{table}

\begin{table}[t]
\centering
\begin{threeparttable}
\caption{Bottom-decile returns: central tendency and minimum-price screens}
\label{tab:central}
\footnotesize
\setlength{\tabcolsep}{3pt}
\begin{tabular}{lccccccc}
\toprule
& \multicolumn{3}{c}{Central tendency (\%/wk)} & \multicolumn{3}{c}{Mean under a minimum-price screen (\%/wk)} & Zero-volume \\
\cmidrule(lr){2-4}\cmidrule(lr){5-7}
Era & Mean & Median & Trim.\ 10--90 & $\ge$\,\$0.000001 & $\ge$\,\$0.0001 & $\ge$\,\$0.01 & share (\%) \\
\midrule
Frontier      & 27.59 & 7.17  & 15.26 & 25.33 & 24.64 & 9.39 & 1.4 \\
Expansion     & 8.05  & $-$2.60 & $-$0.04 & 7.54  & 4.99  & 1.94 & 34.5 \\
Consolidation & 2.79  & $-$0.06 & $-$0.09 & 2.67  & 1.99  & 1.25 & 47.3 \\
\bottomrule
\end{tabular}
\begin{tablenotes}
\footnotesize
\item Notes: Market-adjusted weekly returns of the equal-weighted bottom decile among surviving coins, winsorised cross-sectionally at the weekly 1st/99th percentiles. Columns 5--7 restrict the decile to coins whose quoted price is at least the stated minimum. Weeks are assigned to eras by the snapshot over which the return is realised, whereas Table~\ref{tab:payoff} assigns them by the formation snapshot; the two conventions differ only for weeks straddling an era boundary and move the frontier-era mean from 27.49 to 27.59. ``Median'' and ``Trim.\ 10--90'' compute the within-week cross-sectional median and the mean over the 10th--90th percentile range before averaging across weeks. Columns 5--7 apply a minimum-price screen to bottom-decile constituents before equal-weighting; the median bottom-decile price is \$0.00022 in the frontier era and \$0.00043 in the consolidation era. The final column is the share of bottom-decile coin-weeks reporting zero 24-hour volume, for which the recorded price is stale and the measured return is mechanically zero-then-jump.
\end{tablenotes}
\end{threeparttable}
\end{table}

\begin{table}[t]
\centering
\begin{threeparttable}
\caption{The investable margin: volume screens, timing, turnover, and portfolio breadth}
\label{tab:robust}
\scriptsize
\setlength{\tabcolsep}{4pt}
\begin{tabular}{llcccccc}
\toprule
Volume screen & Timing & Era & D1 adj.\ ret.\ (\%/wk) & ($t$) & Turnover (\%/wk) & Avg.\ \# coins & \# weeks \\
\midrule
$>$\$10k & same-week & Frontier & 2.71 & (0.19) & 46 & 2.0 & 35 \\
$>$\$10k & same-week & Expansion & 6.01$^{***}$ & (2.63) & 42 & 8.9 & 184 \\
$>$\$10k & same-week & Consolidation & 1.97$^{***}$ & (5.45) & 17 & 44.8 & 260 \\
$>$\$10k & skip-week & Frontier & $-$5.31 & ($-$0.49) & 47 & 1.9 & 33 \\
$>$\$10k & skip-week & Expansion & 5.17$^{***}$ & (3.34) & 42 & 8.9 & 183 \\
$>$\$10k & skip-week & Consolidation & 1.12$^{***}$ & (3.23) & 17 & 44.3 & 259 \\
$>$\$100k & same-week & Frontier & --- & --- & 25 & 1.2 & 6 \\
$>$\$100k & same-week & Expansion & 2.66 & (0.48) & 43 & 2.3 & 89 \\
$>$\$100k & same-week & Consolidation & 0.83 & (1.14) & 23 & 12.9 & 255 \\
$>$\$100k & skip-week & Frontier & --- & --- & 25 & 1.2 & 5 \\
$>$\$100k & skip-week & Expansion & 2.00 & (0.45) & 43 & 2.3 & 89 \\
$>$\$100k & skip-week & Consolidation & 1.92$^{***}$ & (3.27) & 22 & 12.7 & 253 \\
$>$\$1m & same-week & Frontier & --- & --- & --- & 1.0 & 1 \\
$>$\$1m & same-week & Expansion & --- & --- & 21 & 1.3 & 19 \\
$>$\$1m & same-week & Consolidation & $-$2.71 & ($-$1.14) & 24 & 1.5 & 97 \\
$>$\$1m & skip-week & Frontier & --- & --- & --- & 1.0 & 1 \\
$>$\$1m & skip-week & Expansion & --- & --- & 17 & 1.3 & 19 \\
$>$\$1m & skip-week & Consolidation & $-$0.30 & ($-$0.12) & 25 & 1.5 & 95 \\
\bottomrule
\end{tabular}
\begin{tablenotes}
\footnotesize
\item Notes: D1 market-adjusted mean weekly return under volume screens of \$10{,}000, \$100{,}000, and \$1{,}000{,}000 (24-hour volume). Same-week: portfolio formed and held over week $t\!\to\!t+1$; skip-week: formed at $t$, held over $t+1\!\to\!t+2$, removing bid--ask bounce. Turnover is the weekly share of portfolio names replaced. Dashes: fewer than 20 tradable weeks. Newey--West $t$-statistics (4 lags). $^{*}$, $^{**}$, $^{***}$: 10\%, 5\%, 1\%.
\end{tablenotes}
\end{threeparttable}
\end{table}

The loosest defensible capacity bound does not change the conclusion. Value-weighting the participation limit across all holdings, rather than binding on the least liquid one, yields \$7,000 in the frontier era, \$80,000 in the expansion era, and \$2.8 million in the consolidation era, against the equal-weighted bounds of \$7,000, \$25,000 and \$293,000 in Table~\ref{tab:capacity}. Even on that reading the strategy that earns 28\% per week on paper could not have absorbed an institutional allocation at any point in the market's history.

\begin{figure}[t]
\centering
\includegraphics[width=\textwidth]{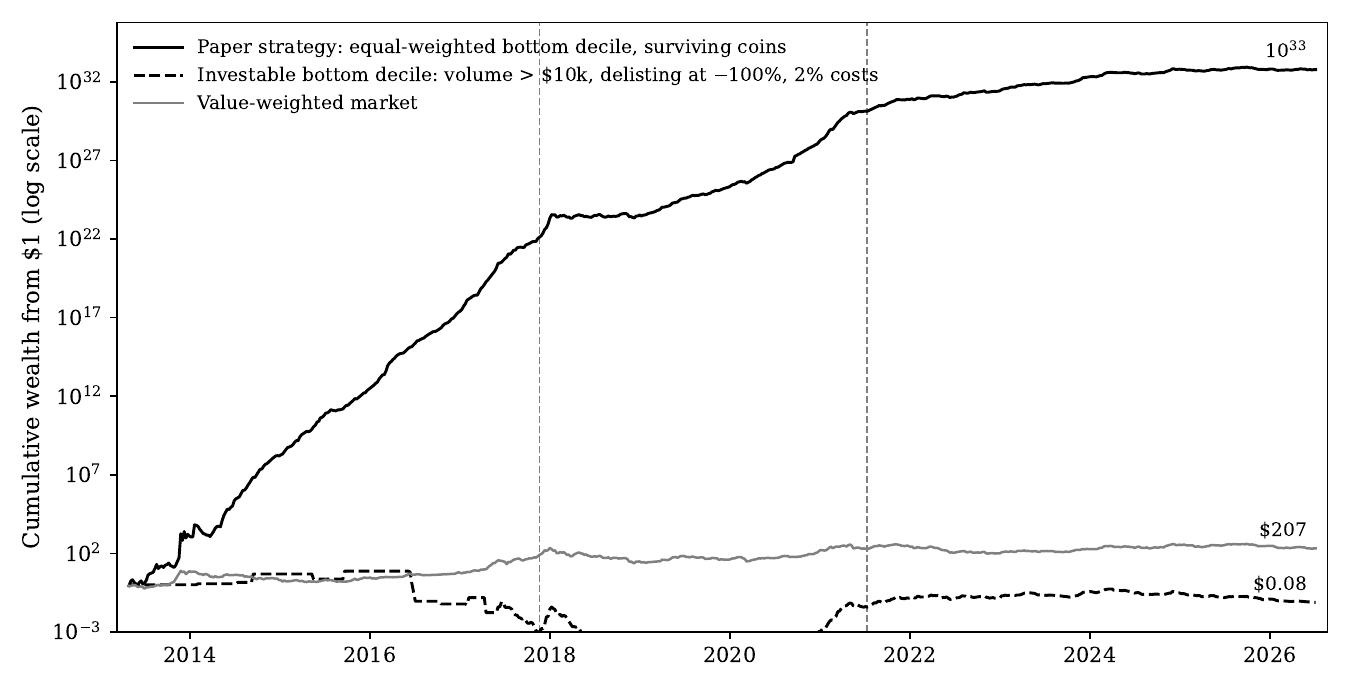}
\caption{Cumulative wealth from \$1 invested in: the paper strategy (equal-weighted bottom decile of surviving coins, weekly rebalancing); the investable bottom decile (volume screen of \$10{,}000, delisting treated as $-100\%$, one-way costs of 2\% on turnover, cash in untradable weeks); and the value-weighted market. Log scale; the three end-of-sample values are annotated. Dashed vertical lines mark the era breaks of 19 November 2017 and 11 July 2021. Weekly data, 28 April 2013 to 12 July 2026; source CoinMarketCap snapshots.}
\label{fig:wealth}

\end{figure}

Figure~\ref{fig:wealth} summarises the argument in one picture: one dollar in the paper strategy compounds to roughly $10^{33}$ dollars over the sample---a magnitude no realisable strategy could produce, and therefore a measure of the distance between the paper portfolio and any implementable one---while the same dollar in the investable version of the strategy, after the volume screen, delisting, and two-percent one-way costs, ends the sample at about eight cents, against roughly \$207 for the value-weighted market.

\subsection{Central tendency, stale prices, and price granularity}
\label{sec:central}

The screens of Table~\ref{tab:robust} ask whether the premium can be captured. Table~\ref{tab:central} asks a prior question: whether the premium describes the typical bottom-decile coin at all. It does not.

Averaged across weeks, the within-week \emph{median} market-adjusted return of a bottom-decile coin is $+7.17\%$ per week in the frontier era but $-2.60\%$ in the expansion era and $-0.06\%$ in the consolidation era. The 10--90 trimmed mean is $+15.26\%$, then $-0.04\%$, then $-0.09\%$. In the two later eras, which together account for 91\% of the sample's coin-weeks, the entire measured premium is generated by the right tail of the cross-section, and the coin at the centre of the bottom decile earns nothing. Winsorising at the 1st and 99th percentiles does not address this: with several hundred to several thousand coins in a weekly cross-section, winsorisation trims a handful of names per side and leaves the tail intact. The sensitivity range reported in Section~\ref{sec:robustness}---23.5\% to 31.8\% for the frontier premium under 2/98 and 0.5/99.5 winsorisation---is itself evidence that the estimate is tail-determined.

Two measurement features drive the tail. The first is price granularity. The median bottom-decile coin is quoted at \$0.00022 in the frontier era and \$0.00043 in the consolidation era, and 39\% and 31\% of bottom-decile coin-weeks respectively are quoted below \$0.0001. At those levels a one-tick move is a large percentage return, and the quoted price carries little information about the price at which any quantity could be transacted. Imposing a minimum price of one cent---still a very low bar---cuts the frontier premium from 27.59\% to 9.39\% per week, the expansion premium from 8.05\% to 1.94\%, and the consolidation premium from 2.79\% to 1.25\%.

The second is staleness. In the consolidation era 47.3\% of bottom-decile coin-weeks report zero 24-hour volume, and 84.2\% report volume below \$10,000. A coin that did not trade carries forward a stale quote, so its measured weekly return is zero until the quote is eventually refreshed, at which point the accumulated move arrives in a single observation. This mechanically produces a distribution with a mass at zero and a long right tail, which is what the median-versus-mean gap in Table~\ref{tab:central} records.

Neither feature is a defect of the data collection; both are accurate descriptions of what a micro-capitalisation cryptocurrency is. But they mean the bottom-decile mean return should not be read as a return an investor could have expected to earn on a typical constituent, and they reinforce the conclusion of Section~\ref{sec:invmargin} from an independent direction. The cumulative-wealth figure of \$$10^{33}$ in Figure~\ref{fig:wealth} is best understood in this light: it is a diagnostic of how far the paper portfolio's measured returns depart from anything realisable, not a return.

\subsection{Factor-adjusted returns}
\label{sec:factors}

Market adjustment may be too coarse a benchmark. Table~\ref{tab:factors} re-estimates the bottom-decile premium as the intercept from time-series regressions on three factors in the spirit of \citet{liu2022common}, constructed internally from the liquid (volume $>$ \$10{,}000) universe: a value-weighted market factor, a size factor (value-weighted bottom minus top capitalisation tertile), and a momentum factor (value-weighted winners minus losers on prior three-week returns); \citet{fieberg2025} document the strength of related trend-based factors in this cross-section. The pattern of Table~\ref{tab:payoff} is unchanged. Among all surviving coins the weekly alpha is 26.6\% in the frontier era, 9.7\% in the expansion era, and 3.0\% in the consolidation era, all overwhelmingly significant, with substantial but insufficient loadings on the size factor. Among liquid coins the frontier-era alpha is statistically zero ($t=0.46$, on 35 weeks and not to be interpreted), while the later-era alphas (5.5\% and 2.0\% per week, gross of costs) mirror the market-adjusted estimates and remain subject to the breadth, turnover, and capacity constraints documented above. The premium is not a repackaged factor loading; it is an untradable residual.

\begin{table}[t]
\centering
\begin{threeparttable}
\caption{Factor-adjusted bottom-decile returns}
\label{tab:factors}
\small
\begin{tabular}{lccccc}
\toprule
Era & $\alpha$ (\%/wk) ($t$) & $\beta_{\mathrm{MKT}}$ & $\beta_{\mathrm{SMB}}$ & $\beta_{\mathrm{MOM}}$ & Weeks \\
\midrule
\multicolumn{6}{l}{\emph{All coins (survivors)}}\\[4pt]
Frontier & 26.57$^{***}$ (8.85) & 0.84 & 0.88 & $-$0.01 & 204 \\
Expansion & 9.68$^{***}$ (8.94) & 1.00 & 0.92 & $-$0.04 & 190 \\
Consolidation & 2.95$^{***}$ (11.21) & 0.88 & 0.53 & 0.04 & 260 \\
\midrule
\multicolumn{6}{l}{\emph{Liquid coins (vol.\ $>$ \$10k)}}\\[4pt]
Frontier & 6.44 (0.46) & $-$1.19 & 1.01 & 0.70 & 35 \\
Expansion & 5.51$^{***}$ (2.72) & 0.97 & 1.63 & $-$0.31 & 184 \\
Consolidation & 2.02$^{***}$ (5.02) & 0.86 & 0.83 & 0.04 & 260 \\
\bottomrule
\end{tabular}
\begin{tablenotes}
\footnotesize
\item Notes: Frontier-era regressions use 204 of the era's 239 weeks because the factors require three weeks of prior returns and 24-hour volume, which is unavailable in the 2013 snapshots. The frontier-era liquid row rests on 35 weekly observations and should not be interpreted as an estimate: its market beta of $-1.19$ is not a meaningful quantity for a cryptocurrency portfolio, and reflects the near-absence of a tradable bottom decile in that era documented in Table~\ref{tab:robust}. Intercepts and loadings from time-series regressions of the equal-weighted bottom-decile portfolio return on three factors constructed from the liquid universe (24h volume $>$ \$10{,}000): value-weighted market; size (value-weighted bottom-minus-top capitalisation tertile); momentum (value-weighted winners-minus-losers on prior 3-week returns, 30/70 breakpoints). Newey--West $t$-statistics (4 lags). $^{*}$, $^{**}$, $^{***}$: 10\%, 5\%, 1\%.
\end{tablenotes}
\end{threeparttable}
\end{table}

\subsection{Testing payoff antisymmetry}
\label{sec:payoffsym}

The mapping from transition symmetry to efficiency in Section~\ref{sec:theory} maintains payoff antisymmetry: $\mu_{i,i+k}=-\mu_{i,i-k}$. Table~\ref{tab:paysym} tests it directly, comparing the mean realised return of coins moving up $k$ deciles with that of coins moving down $k$ deciles. Antisymmetry is firmly rejected: up-moves pay considerably more than down-moves cost ($\mu^{U}_{k}+\mu^{D}_{k}>0$ for every $k$ and era), reflecting the mechanical convexity of the rank--return relation---moving up a decile requires outgrowing a boundary, and the return distribution of micro-capitalisation coins is strongly right-skewed. As Section~\ref{sec:theory} explains, this is what the return distribution guarantees rather than a substantive economic finding---a cross-section symmetric in logs and free of any predictability produces $\mu^U_k+\mu^D_k>0$, rising in $k$, of the same order of magnitude as the estimates here. The table should therefore be read as measuring the size of the wedge between mobility symmetry and return efficiency, not as testing a hypothesis that might plausibly have held. Two implications follow. First, the two asymmetries partially offset: measured transition probabilities tilt downward for survivors (Table~\ref{tab:zstats}) while payoffs tilt upward, which is why realised decile portfolio returns (Table~\ref{tab:payoff}), the economically meaningful object, are the appropriate summary rather than either asymmetry alone. Second, the payoff skew itself compresses sharply across eras---the $k=1$ asymmetry falls from 36.8\% per week in the frontier era to 14.8\% in the consolidation era---so both components of the efficiency violation converge as the market matures.

\begin{table}[t]
\centering
\begin{threeparttable}
\caption{Payoff antisymmetry: realised returns of up- versus down-movers}
\label{tab:paysym}
\small
\begin{tabular}{lccccc}
\toprule
Era & $k$ & $\mu^{U}_{k}$ (\%) & $\mu^{D}_{k}$ (\%) & $\mu^{U}_{k}+\mu^{D}_{k}$ (\%) & ($t$) \\
\midrule
Frontier & 1 & 63.7 & $-$28.5 & 36.8$^{***}$ & (4.57) \\
 & 2 & 200.9 & $-$60.2 & 147.9$^{***}$ & (16.17) \\
 & 3 & 243.6 & $-$70.0 & 192.4$^{***}$ & (10.71) \\
Expansion & 1 & 54.7 & $-$26.1 & 28.6$^{***}$ & (6.47) \\
 & 2 & 164.8 & $-$58.4 & 108.4$^{***}$ & (9.24) \\
 & 3 & 192.0 & $-$61.4 & 139.3$^{***}$ & (6.27) \\
Consolidation & 1 & 36.0 & $-$21.2 & 14.8$^{***}$ & (7.43) \\
 & 2 & 84.6 & $-$48.0 & 36.4$^{***}$ & (8.92) \\
 & 3 & 85.1 & $-$49.0 & 33.2$^{***}$ & (5.15) \\
\bottomrule
\end{tabular}
\begin{tablenotes}
\footnotesize
\item Notes: $\mu^{U}_{k}$ ($\mu^{D}_{k}$) is the mean weekly winsorised return of coins moving up (down) exactly $k$ deciles, averaged within weeks and then across weeks. Under payoff antisymmetry $\mu^{U}_{k}+\mu^{D}_{k}=0$. Newey--West $t$-statistics (4 lags) on the weekly series of $\mu^{U}_{k,t}+\mu^{D}_{k,t}$. $^{*}$, $^{**}$, $^{***}$: 10\%, 5\%, 1\%.
\end{tablenotes}
\end{threeparttable}
\end{table}

Tables~\ref{tab:payoff}--\ref{tab:paysym} and Figure~\ref{fig:wealth} together establish the paper's payoff conclusion. The cryptocurrency size premium is real as a statistical object and enormous on paper; it was never a portfolio strategy. In the frontier era it existed only among untradable coins. In the consolidation era it survives a \$10,000 screen, and survives costs, but only when deciles are formed on the whole universe: re-formed inside the tradable universe the gradient is $0.01\%$ per week ($t=0.04$), and the portfolio that carries the screened premium holds a few hundred thousand dollars. The expansion era is the partial exception, retaining a genuine tradable gradient of $-2.32\%$ per week. Across the market's history, and unambiguously in its modern phase, the statistical inefficiency has been confined to the segment that cannot absorb capital---an illiquidity mirage.

\section{Attention, survival, and the structure that remains}
\label{sec:framework}

\subsection{Motivation}
\label{sec:fw:motivation}

Sections~\ref{sec:symmetry} and~\ref{sec:payoff} establish two central findings. First, the cryptocurrency market exhibits persistent departures from Markov mobility efficiency, although the magnitude of those departures declines substantially over time. Second, much of the associated return predictability disappears once delisting risk, liquidity constraints, turnover costs, and limited portfolio capacity are considered. The remaining question is therefore not whether asymmetry exists, but what economic forces organise the cross-sectional structure that survives as the market matures.

A frictionless account does not fit these patterns, but neither does the familiar conclusion that the cryptocurrency size premium partly reflects compensation for illiquidity \citep{liu2022common}. Three features of our evidence go further. First, among tradable assets in the consolidation era there is no size gradient to compensate: re-forming deciles within the \$10,000-volume universe gives a spread between the largest and smallest deciles of 0.01\% per week, and at the \$1,000,000 screen returns rise weakly with size. Second, outside the frontier era the premium is not earned by the typical small coin, whose median return is negative; it is generated by a right tail of sub-cent, zero-volume coins whose returns arrive as stale-price jumps. Third, the decline in mobility asymmetry is compositional, driven by the entry rate, rather than a repricing of incumbents. What remains to be explained is the persistence and survival structure that is left once the return gradient has gone. That structure describes an environment in which information acquisition is costly, investor attention is unevenly distributed, and survival itself is uncertain.

To rationalise these patterns, we propose a parsimonious attention-and-survival framework built on three ideas. First, investor attention is scarce. Because acquiring and processing information is costly, investors cannot monitor the full universe of cryptocurrencies and must allocate attention selectively. Second, attention varies systematically across the capitalisation distribution. Large-capitalisation assets attract greater media coverage, analyst attention, exchange support, and institutional participation, whereas smaller assets are more likely to remain neglected. Following the investor-recognition hypothesis of \citet{merton1987}, investor recognition, denoted $a_i$, is interpreted as the proportion of investors who are aware of and actively consider the assets in capitalisation decile $i$ for investment. Throughout the remainder of the paper, references to investor attention should be understood as shorthand for investor recognition. Third, exposure to neglect and survival risk may command compensation in equilibrium. Assets receiving less attention and facing a greater probability of economic failure must offer higher expected returns to attract capital.

The central mechanism linking these ideas is investor recognition. Investors form expectations using the information they choose to process, so recognition affects both beliefs and capital allocation. Greater recognition expands the effective investor base of an asset, while exposure to catastrophic outcomes---delisting, project abandonment, exchange removal, or permanent destruction of value---discourages participation. Letting $Q_i$ denote demand for the assets in decile $i$ and $C_i$ their exposure to catastrophic survival risk, the mechanism requires
\begin{equation}
\frac{\partial Q_i}{\partial a_i} > 0,
\qquad
\frac{\partial Q_i}{\partial C_i} < 0.
\label{eq:fw:demand}
\end{equation}
Greater recognition therefore supports liquidity, exchange-listing continuity, developer participation, and capital commitment. Assets receiving broad recognition benefit from larger and potentially more resilient investor bases, whereas neglected assets depend on narrower pools of capital and face weaker prospects for continued economic survival.

These demand effects provide the common foundation for the framework's empirical predictions, and it is worth stating them as maintained relationships before deriving the restrictions they imply. First, a broader and more resilient investor base dampens the effect of transitory idiosyncratic shocks on an asset's relative market position, so that the probability $p_{ii}$ of remaining in the current capitalisation state rises with recognition. Second, sustained investor participation, liquidity, exchange support, and ecosystem activity reduce the likelihood that an asset disappears from the observed market, so the exit hazard $h_i$ falls with recognition. Third, investors must be compensated for holding assets that attract narrow investor bases, receive limited information production, and face an elevated probability of economic failure, so conditional expected returns $\mu_i$---defined formally in Equation~\eqref{eq:fw:mudef}---rise with exit risk and fall with recognition. Compactly,
\begin{equation}
\frac{\partial p_{ii}}{\partial a_i} > 0,
\qquad
\frac{\partial h_i}{\partial a_i} < 0,
\qquad
\frac{\partial \mu_i}{\partial h_i} > 0,
\qquad
\frac{\partial \mu_i}{\partial a_i} < 0.
\label{eq:fw:mech}
\end{equation}
Predictions 1--3 below are consequences of the maintained relationships in Equations~\eqref{eq:fw:demand} and~\eqref{eq:fw:mech} together with the premise that recognition rises with market capitalisation, rather than three separately postulated hypotheses. The first relation in~\eqref{eq:fw:mech} concerns positional \emph{stability} and not the \emph{direction} of movement: the framework predicts higher staying probabilities among more widely recognised assets, not systematically greater upward mobility. This distinction matters because one-step directional transition intensities may be dominated by the greater return volatility of smaller cryptocurrencies, while measured persistence in the lower tail may be affected by exit truncation and the bounded nature of the ranking. Section~\ref{sec:cck} shows that both of these boundary forces are present in the data.

The mechanism does not depend on behavioural biases or investor irrationality. Limited attention arises as a rational response to information-processing constraints in a market where monitoring the full asset universe is prohibitively costly. Investor recognition is therefore not treated as an exogenous behavioural preference. It is an equilibrium state variable associated with the size and resilience of an asset's investor base. Capitalisation, recognition, liquidity, exchange support, project quality, survival prospects, and expected returns may all be jointly determined.

The framework does not identify a unique structural attention process, nor the causal effect of investor recognition on market outcomes. Instead, it identifies a set of qualitative implications that must hold under the maintained relationships linking recognition, demand, persistence, survival, and compensation. Because it generates directional rather than point predictions, its empirical content resides in the ordering of observable outcomes across the capitalisation distribution: the appropriate question is not whether a particular structural coefficient equals a specific value, but whether the collection of model-implied cross-sectional inequalities is satisfied individually and jointly. The following subsections derive three such cross-sectional restrictions---governing positional persistence, survival risk, and conditional compensation---together with a complementary market-development implication that follows from the same premise about recognition and market breadth but is evaluated on a different unit of observation.

\subsection{Attention and persistence}
\label{sec:fw:persistence}

The first implication of scarce attention is selective coverage. Cryptocurrency markets contain thousands of competing assets, yet investors possess limited information-processing capacity and cannot monitor every project simultaneously. Larger-capitalisation assets therefore attract a disproportionate share of aggregate attention.

Let
\[
s_i \;\equiv\; p_{ii} \;=\; \Pr\!\left(S_{t+1}=i \mid S_t=i,\; S_{t+1}\neq X\right)
\]
denote the probability that a surviving asset remains in capitalisation decile $i$ from one week to the next---the staying probability of Section~\ref{sec:risk}, and the object signed by the first relation of Equation~\eqref{eq:fw:mech}. Its expected sojourn duration is
\begin{equation}
\mathbb{D}_i \;=\; \frac{1}{1-s_i}.
\label{eq:fw:sojourn}
\end{equation}

Greater attention can stabilise an asset's position within the cross-sectional ranking. Widely followed assets benefit from larger and potentially more resilient investor bases, making their relative standing less sensitive to transitory idiosyncratic shocks. This is the content of $\partial p_{ii}/\partial a_i>0$ in Equation~\eqref{eq:fw:mech}, and it bears on stability alone.

For the part of the distribution over which attention is the operative margin, the framework implies
\begin{equation}
s_{k^{*}} \;\le\; s_{k^{*}+1} \;\le\; \cdots \;\le\; s_{K},
\qquad\text{equivalently}\qquad
\mathbb{D}_{k^{*}} \;\le\; \cdots \;\le\; \mathbb{D}_{K},
\label{eq:fw:pers}
\end{equation}
where $K$ is the top decile as defined in Section~\ref{sec:theory} and $k^{*}$ identifies the lowest capitalisation decile for which recognition-related differences are expected to dominate truncation effects.

The restriction deliberately excludes the lower tail of the distribution. Among the smallest assets, delisting and project failure remove poorly performing cryptocurrencies from the transition sample, while the bounded nature of the ranking mechanically limits further downward movement. These truncation effects may elevate observed persistence independently of investor recognition. Consequently, the framework predicts that recognition becomes the dominant determinant of positional stability only beyond a threshold level of market capitalisation, denoted $k^{*}$. Because the objective is to derive qualitative restrictions rather than to estimate a structural attention process, $k^{*}$ should be interpreted as the point beyond which differences in investor recognition are expected to dominate truncation effects.

The framework does not deliver the numerical value of $k^{*}$, so it must be fixed empirically, and we do so using a statistic that is distinct from the restriction being tested and estimated in the earliest part of the sample. Because truncation compresses downward movement in the lower tail while recognition stabilises positions in the upper tail, the profile of staying probabilities across deciles is U-shaped, and the decile at which it turns is a natural estimate of the boundary between the two regimes. In the frontier era the profile runs $0.847, 0.676, 0.633, 0.629, 0.643, \dots$, so the turning point is decile~4, and the expansion era independently agrees; in the consolidation era deciles~4 and~5 are identical to three decimal places. We therefore set $k^{*}=4$ on the basis of the frontier era and apply it unchanged to the two later eras, which serve as out-of-sample validation. Section~\ref{sec:robustness} reports the complete sensitivity profile over $k^{*}$.

\begin{quote}
\noindent\textbf{Prediction 1 (Persistence).} Higher market capitalisation attracts greater investor attention, which enhances positional stability and reduces sensitivity to transitory shocks. Consequently, among assets in the attention-differentiated upper segment of the distribution ($i \ge k^{*}$), larger cryptocurrencies are expected to exhibit higher staying probabilities and longer expected sojourn durations.
\end{quote}

\subsection{Neglect and survival risk}
\label{sec:fw:survival}

The second implication of the framework concerns economic survival. Attention affects not only positional stability but also the probability that an asset continues to exist. Because investor attention is finite, the share of investors who recognise, monitor, and actively consider an asset varies systematically across the capitalisation distribution, with larger assets attracting a disproportionately larger share of aggregate attention. Conversely, assets receiving relatively little attention face a higher probability of delisting, abandonment, or economic disappearance.

Define the exit hazard from decile $i$ as
\begin{equation}
h_i \;=\; \Pr\!\left(S_{t+1}=X \mid S_t=i\right).
\label{eq:fw:hazdef}
\end{equation}
If attention increases with market capitalisation and neglect increases the probability of economic disappearance, the framework implies
\begin{equation}
h_1 \;>\; h_2 \;>\; \cdots \;>\; h_K.
\label{eq:fw:haz}
\end{equation}

Unlike conventional equities, cryptocurrency projects can disappear abruptly when trading activity collapses, exchanges withdraw support, or project development ceases. Survival risk is therefore a first-order dimension of the cryptocurrency cross-section.

Equation~\eqref{eq:fw:haz} should not be interpreted as a causal model of delisting. Market capitalisation, attention, liquidity, exchange support, project quality, and survival prospects are jointly determined in equilibrium. Instead, the restriction represents a qualitative implication of the framework: if attention is concentrated among larger assets and survival becomes more difficult as attention declines, economic disappearance should be concentrated among the smallest cryptocurrencies.

\begin{quote}
\noindent\textbf{Prediction 2 (Survival risk).} Exit hazards decrease monotonically with market capitalisation, as larger assets benefit from greater investor attention, stronger ecosystem support, and enhanced prospects for continued survival.
\end{quote}

Survival risk matters not only because it affects market participation, but also because investors may require compensation for bearing it.

\subsection{Conditional compensation for neglect}
\label{sec:fw:returns}

The third implication of the framework concerns expected returns. Investors holding small cryptocurrencies are exposed to risks that conventional measures of return volatility may not fully capture. Limited attention is associated with lower liquidity, weaker ecosystem support, greater informational opacity, and a heightened probability of economic failure. In an equilibrium characterised by scarce attention and heterogeneous neglect, investors may therefore require compensation for bearing both neglect and survival risk.

One notational point before proceeding. Section~\ref{sec:theory} writes $\mu_{ij}$ for the payoff attached to a \emph{transition} from decile $i$ to decile $j$; the single-subscript $\mu_i$ used here and in Prediction~3 is a different object, the expected return of a decile conditional on survival, and both are total returns rather than market-adjusted ones. The empirical counterparts in Sections~\ref{sec:payoff} and~\ref{sec:drivers} are market-adjusted; because the market return is common across deciles within a week, the ordering the prediction imposes is invariant to that adjustment. Define $\mu_i$ as the expected return of assets beginning in capitalisation decile $i$, conditional on remaining in the observed market through period $t+1$:
\begin{equation}
\mu_i \;=\; E\!\left(R_{t+1} \mid S_t=i,\; S_{t+1}\neq X\right).
\label{eq:fw:mudef}
\end{equation}
Recalling that $a_i$ denotes the recognition received by assets in decile $i$ and $h_i$ their exit hazard, a parsimonious representation of conditional compensation is
\begin{equation}
\mu_i \;=\; r_f \;+\; \phi_1\,\mathrm{Var}(R_i) \;+\; \phi_2\,h_i \;+\; \phi_3\,(1-a_i),
\qquad \phi_1,\phi_2,\phi_3>0.
\label{eq:fw:mu}
\end{equation}
The variance term is the conventional risk--return channel and plays no role in the framework's comparative statics; we sign it positive for completeness and do not use it. The positive association with $h_i$ captures ex ante compensation for bearing elevated failure risk, observed ex post among the subset of assets that survives. The negative association with $a_i$ captures compensation for neglect: smaller cryptocurrencies receive less attention and face greater uncertainty regarding long-run viability, and may therefore command higher expected returns conditional on survival.

If attention increases and exit risk decreases with market capitalisation, the comparative statics imply the testable size ordering
\begin{equation}
\mu_1 \;>\; \mu_2 \;>\; \cdots \;>\; \mu_K.
\label{eq:fw:ret}
\end{equation}
This ordering constitutes the central empirical prediction of the return component of the framework.

\begin{quote}
\noindent\textbf{Prediction 3 (Conditional compensation).} Because smaller assets are exposed to greater neglect and survival risk, surviving assets from lower-capitalisation deciles are expected to earn higher returns. Consequently, conditional expected returns decline monotonically across the capitalisation distribution of the capitalisation-reporting population; see the scope condition immediately below.
\end{quote}

Prediction~3 carries a scope condition that the other two do not. Recognition is what the mechanism prices, and an asset recognised by few investors is an asset that few investors trade. The compensation the prediction describes is therefore expected precisely where trading activity is lowest, and there is no reason internal to the framework to expect it among assets that trade freely: an asset with deep and continuous two-sided volume is, by the framework's own definition, not neglected, whatever its capitalisation. The condition bites where recognition and tradability have separated. Where they have not---as in the expansion era, when even the traded segment was thinly followed---a size gradient among tradable assets is entirely consistent with the mechanism, and Section~\ref{sec:invmargin} finds one. Section~\ref{sec:liquid} tests the prediction on progressively more liquid universes and finds exactly this---the ordering holds on the full population, and is rejected in the consolidation era once a trading-activity screen is imposed.

Equations~\eqref{eq:fw:mu} and~\eqref{eq:fw:ret} are maintained comparative-static restrictions rather than structurally identified causal effects. As noted above, these quantities are jointly determined in equilibrium. The empirical analysis therefore evaluates whether the data exhibit the size ordering implied by Equation~\eqref{eq:fw:ret}, rather than attempting to identify an independent causal effect of latent attention on expected returns.

Moreover, the survival conditioning in Equation~\eqref{eq:fw:mudef} is economically important. Higher returns among surviving small-capitalisation assets do not imply that investors realise comparable returns across the full population of cryptocurrencies. Realised exits may eliminate part of the apparent compensation, while limited liquidity, high turnover, transaction costs, and portfolio-capacity constraints may absorb much of the remainder. Section~\ref{sec:payoff} documents precisely this distinction. The framework therefore interprets the observed size premium as conditional compensation for bearing neglect and survival risk within a difficult investment environment, rather than as freely available abnormal performance.

\subsection{Market development and the compression of asymmetry}
\label{sec:fw:development}

The framework further implies that the cross-sectional consequences of attention should weaken as market breadth expands and information becomes more widely dispersed across the asset universe.

Section~\ref{sec:owncap} shows that the principal converging component of directional asymmetry is compositional. A high entry rate during the market's frontier phase mechanically displaced incumbent assets through the relative ranking. As entry slowed relative to the installed asset base, this source of rank churn weakened. By contrast, asymmetry measured against frozen decile boundaries, which isolates changes in coins' own capitalisations, remains comparatively stable across eras.

The interpretation of convergence therefore distinguishes two related mechanisms. First, declining entry-driven compositional churn directly reduces directional asymmetry in relative ranks. Second, market deepening may reduce informational frictions and attenuate the distinction between highly attended assets and the neglected fringe. The first mechanism is established by the decomposition in Section~\ref{sec:owncap}. The second provides a complementary equilibrium interpretation but is not separately identified as the causal source of convergence.

Let $\mathcal{A}_t$ denote the directional asymmetry index introduced in Section~\ref{sec:symmetry}, $\mathcal{M}_t$ a measure of market maturity such as market breadth, and $\mathcal{E}_t$ the rate of new entry. The empirical relationship can be expressed as
\begin{equation}
\mathcal{A}_t \;=\; f\!\left(\mathcal{M}_t,\, \mathcal{E}_t\right),
\qquad
\frac{\partial f}{\partial \mathcal{M}} < 0,
\qquad
\frac{\partial f}{\partial \mathcal{E}} > 0.
\label{eq:fw:dev}
\end{equation}
The first derivative captures the prediction that asymmetry is lower in broader and more developed market states. The second captures the compositional prediction that a larger entry margin generates greater relative-rank displacement.

Equation~\eqref{eq:fw:dev} represents a conditional empirical relationship rather than a structural causal model. Market breadth and entry intensity may be closely related, and maturity simultaneously captures changes in the installed asset base, liquidity, information availability, institutional participation, and market composition. Section~\ref{sec:drivers} therefore evaluates the two dimensions using complementary evidence. The frozen-boundary decomposition identifies the importance of entry-driven displacement, while the time-series analysis examines whether broader market states are associated with lower asymmetry.

\begin{quote}
\noindent\textbf{Prediction 4 (Market development).} If directional asymmetry originates from limited attention, rapid entry, and an underdeveloped information environment, its magnitude should diminish as the market broadens and matures. Consequently, directional asymmetry is predicted to decrease with market breadth, increase with entry intensity, and converge gradually rather than through isolated institutional events.
\end{quote}

\subsection{Scope and empirical content}
\label{sec:fw:scope}

The attention-and-survival framework maps the empirical regularities documented in Sections~\ref{sec:symmetry} and~\ref{sec:payoff} into four testable restrictions governing market development, positional persistence, survival risk, and conditional compensation. These restrictions are intentionally confined to the margins the framework is designed to explain. In particular, the persistence prediction applies only beyond the threshold $k^{*}$, where differences in investor recognition are expected to dominate the truncation and boundary effects that characterise the lower tail of the capitalisation distribution.

The framework's empirical content derives not from its ability to rationalise the observed hierarchy ex post, but from the requirement that its cross-sectional restrictions hold jointly in the data. We state what would count against it. A violation of the survival ordering across the capitalisation distribution would be inconsistent with the framework's central implications on any universe. A violation of the persistence ordering above $k^{*}$ would be inconsistent with them on any universe whose deciles are re-formed within it; we exclude universes formed by screening a full-universe sort, because that operation empties the lower deciles unevenly and leaves their staying probabilities estimated on a small and selected remnant rather than on a decile. A violation of the conditional-compensation ordering would be inconsistent with the framework \emph{on the capitalisation-reporting population}; it does not claim the ordering on tradable subsets, for the reason set out below, and Section~\ref{sec:invmargin} establishes that the tradable cross-section is a different object before the framework is written down. Prediction~4, which is not part of the joint test, would be contradicted by a directional asymmetry that failed to decline as the entry margin contracted, or by a frozen-boundary component that declined as fast as the relative-rank component; the decomposition of Section~\ref{sec:owncap} is what puts it at risk.

The predictions can be expressed compactly as collections of ordered restrictions. For prediction $j$, let $m^{(j)}$ denote the vector of moment functions whose expectations the prediction signs, so that
\begin{equation}
H_0^{(j)}:\quad E\!\left[m^{(j)}_k\right] \le 0 \quad \text{for every element } k,
\label{eq:fw:hj}
\end{equation}
with $m^{(1)}_k = s_k - s_{k+1}$ for $k\ge k^{*}$ (persistence), $m^{(2)}_k = h_{k+1}-h_k$ (survival risk), and $m^{(3)}_k = \mu_{k+1}-\mu_k$ (conditional compensation). Predictions 1--3 are cross-sectional, are stated over the same decile index, and share the recognition-demand foundation of Equations~\eqref{eq:fw:demand} and~\eqref{eq:fw:mech}. They therefore constitute a single system, and the appropriate null is their intersection,
\begin{equation}
H_0:\quad \bigcap_{j=1}^{3} H_0^{(j)},
\label{eq:fw:joint}
\end{equation}
tested as one family of many moment inequalities in Section~\ref{sec:cck}.

Prediction~4 is deliberately excluded from~\eqref{eq:fw:joint}. It is a time-series statement about the market's aggregate development rather than a cross-sectional ordering across deciles, it is evaluated on a different unit of observation, and one component of it---the role of the entry margin---is established by the frozen-boundary decomposition rather than estimated as a signed derivative. It is therefore a complementary market-development implication of the framework, evaluated in Section~\ref{sec:drivers} through the decomposition, the time-series regression, and the event studies, and it should not be read as entering the formal joint test.

A second scope condition concerns the universe over which the restrictions are imposed, and unlike the threshold $k^{*}$ it is not a modelling convenience but a consequence of the mechanism. Equation~\eqref{eq:fw:demand} makes recognition the driver of demand, and demand is what produces trading volume; recognition and tradability are therefore two readings of one latent quantity, not independent regressors. An asset neglected enough to command a recognition premium is, very nearly by construction, an asset too neglected to trade in size. It follows that imposing a volume screen does not hold the mechanism fixed while varying liquidity---it removes the very assets on which the compensation channel operates. The persistence and survival channels carry no such implication, because Equation~\eqref{eq:fw:mech} signs them through positional stability and ecosystem support rather than through the marginal investor's willingness to hold an untraded asset; they should and do survive screening. Section~\ref{sec:invmargin} shows independently, and without reference to the framework, that the size gradient among tradable assets in the consolidation era is $0.01\%$ per week. The system in~\eqref{eq:fw:joint} should accordingly be understood as a description of the capitalisation-reporting population, not of the subset of it that supports meaningful trading.

Failure to reject~\eqref{eq:fw:joint} would indicate that the residual cross-sectional structure is consistent with the attention-and-survival framework. Such evidence would not establish recognition as a uniquely identified causal mechanism, since recognition, market capitalisation, liquidity, project quality, and survival prospects are jointly determined in equilibrium. Nor, being the failure of a test to reject, would it establish that every inequality holds. Rather, it would show that the observed patterns of mobility, survival, and conditional compensation lie inside the identified set implied by a cross-section shaped by differences in recognition and survival risk. Because the restrictions are inequalities rather than point conditions, the appropriate empirical framework is a many-moment-inequality test \citep{cck2014}; identification is deliberately partial in the sense of \citet{molinari2020}, trading parametric precision for robustness of the qualitative predictions.

Section~\ref{sec:drivers} evaluates Predictions~1--3 individually and jointly, and Prediction~4 separately. Collectively, they transform the empirical findings from a collection of stylised facts into a coherent account linking persistence, survival, conditional compensation, and market development through a common recognition-based mechanism.

\FloatBarrier

\section{Empirical evaluation of the attention-and-survival framework}
\label{sec:drivers}

The framework developed in Section~\ref{sec:framework} provides an economic interpretation for the residual asymmetries that survive as the cryptocurrency market converges towards Markov mobility efficiency. Rather than seeking to explain every feature of cross-sectional mobility, it characterises the structure that remains through systematic differences in visibility, persistence, survival risk, and conditional return compensation across the capitalisation distribution. It delivers four empirical predictions: directional asymmetry should be lower in more mature market states while increasing with entry-driven compositional churn (Prediction~4); persistence should increase within the attention-differentiated upper segment of the distribution, $i\ge k^{*}$ (Prediction~1); exit hazards should decline with size (Prediction~2); and conditional survivor returns should decline with size (Prediction~3). We take them in the order in which the supporting evidence accumulates---first the time-series and event-study evidence bearing on market development, then the cross-sectional evidence on persistence, survival, and compensation---before assessing all of the cross-sectional restrictions jointly using formal many-inequality tests.

\subsection{What compresses the asymmetry? (Prediction 4)}

Equation~\eqref{eq:fw:dev} predicts that asymmetry falls with market maturity and rises with entry intensity. Both legs can be taken to the time series, but doing so credibly requires more care than a single regression affords, because market breadth is very close to a deterministic trend over this sample: log breadth correlates 0.86 with calendar time, and $-0.96$ with the entry rate when both are averaged annually. A regression of a smoothed dependent variable on a trending regressor will produce a large $R^2$ and a large $t$-statistic whether or not any economic relationship exists. Table~\ref{tab:drivers} therefore reports the breadth and entry margins separately and jointly, under three inference schemes of increasing conservatism.

The results do not support the claim that breadth is the robust correlate. In weekly data, using the unsmoothed asymmetry index rather than its 52-week moving average, breadth enters with a coefficient of $-0.053$ ($t=-8.2$) and entry with $+1.36$ ($t=5.4$); in the joint specification breadth dominates and entry is insignificant. In annual non-overlapping averages---thirteen genuinely independent observations, and the specification least contaminated by the overlapping-window problem---the ranking reverses completely: entry alone explains 98\% of the variation ($t=19.4$), and in the joint specification entry retains a coefficient of $+3.49$ ($t=8.6$) while breadth changes sign and becomes insignificant ($t=1.5$). In first differences, which strip both trends, breadth is indistinguishable from zero in every specification, while entry enters with the wrong sign---significantly so on its own ($t=-2.63$) and marginally in the joint specification ($t=-1.66$).

We draw three conclusions. First, the association between asymmetry and the state of the market is real: it survives removing the moving-average smoothing, which was the specification most open to the charge of manufacturing significance. Second, breadth and entry cannot be separated at the frequency at which the convergence question is posed; with an annual correlation of $-0.96$ the joint specifications are not identified, and which regressor wins is a function of the frequency examined rather than of the economics. Third, to the extent that the two can be told apart at all, the entry margin has at least as strong a claim as breadth, which is what the frozen-boundary decomposition of Section~\ref{sec:owncap} independently implies: what converged is entry-driven compositional churn, not the informational environment. We therefore treat Prediction~4 as supported in its compositional leg, established by the decomposition rather than by this regression, and as not separately identified in its maturity leg. Breadth is not the robust correlate of asymmetry once inference accounts for the overlapping windows.

These are equilibrium correlations, not causal estimates. The event-study evidence next probes whether discrete institutional milestones moved the process.

\begin{table}[t]
\centering
\begin{threeparttable}
\caption{Directional asymmetry, market breadth, and the entry margin}
\label{tab:drivers}
\small
\setlength{\tabcolsep}{5pt}
\begin{tabular}{lcccccc}
\toprule
& \multicolumn{2}{c}{Weekly, unsmoothed} & \multicolumn{2}{c}{First differences} & \multicolumn{2}{c}{Annual, non-overlapping} \\
\cmidrule(lr){2-3}\cmidrule(lr){4-5}\cmidrule(lr){6-7}
& Coefficient & ($t$) & Coefficient & ($t$) & Coefficient & ($t$) \\
\midrule
\multicolumn{7}{l}{\emph{Panel A: market breadth only}}\\
$\log$(number of coins) & $-$0.0525 & ($-$8.17) & $-$0.2168 & ($-$0.49) & $-$0.0507 & ($-$5.92) \\
$R^2$ & \multicolumn{2}{c}{0.461} & \multicolumn{2}{c}{0.008} & \multicolumn{2}{c}{0.880} \\
\midrule
\multicolumn{7}{l}{\emph{Panel B: entry rate only}}\\
Entry rate & 1.3595 & (5.40) & $-$0.7616 & ($-$2.63) & 2.9378 & (19.44) \\
$R^2$ & \multicolumn{2}{c}{0.240} & \multicolumn{2}{c}{0.078} & \multicolumn{2}{c}{0.977} \\
\midrule
\multicolumn{7}{l}{\emph{Panel C: both}}\\
$\log$(number of coins) & $-$0.0472 & ($-$6.82) & 0.0285 & (0.06) & 0.0105 & (1.49) \\
Entry rate & 0.3060 & (0.86) & $-$0.7737 & ($-$1.66) & 3.4947 & (8.63) \\
$R^2$ & \multicolumn{2}{c}{0.468} & \multicolumn{2}{c}{0.078} & \multicolumn{2}{c}{0.979} \\
\midrule
Observations & \multicolumn{2}{c}{688} & \multicolumn{2}{c}{687} & \multicolumn{2}{c}{13} \\
\bottomrule
\end{tabular}
\begin{tablenotes}
\footnotesize
\item Notes: Dependent variable is the weekly directional asymmetry index $\mathcal{A}$ of equation~\eqref{eq:asym}, computed from that week's transition counts, not smoothed. Columns 1--2 use the weekly series with Newey--West standard errors (8 lags); columns 3--4 first-difference every series and use the same standard errors; columns 5--6 average all series within thirteen non-overlapping 52-week blocks and estimate by OLS, so that no observation shares data with another. The entry rate is the share of a week's cross-section not present in the preceding week. Correlations that bear on identification: $\log$(breadth) with calendar time, 0.86; entry rate with calendar time, $-0.47$; breadth with entry, $-0.62$ weekly and $-0.96$ in annual averages.
\end{tablenotes}
\end{threeparttable}
\end{table}

\subsection{Spot-ETF approvals: no event-time discontinuity (Prediction 4)}
\label{sec:event}

The approval of spot Bitcoin ETFs in January 2024 is the sharpest institutional milestone in the sample, and the launch of spot Ether ETFs in July 2024 a second one. If institutional access per se disciplines the cross-section, the mobility of institutionally exposed coins should compress discretely around these dates. Table~\ref{tab:event} implements a difference-in-differences-style event comparison. The outcome is the weekly probability of moving deciles (conditional on survival) for treated coins (deciles 9--10, where ETF flows and their spillovers concentrate) minus control coins (deciles 4--6), compared across $\pm 26$-week windows around each event, with Newey--West inference on the weekly difference series and January 2023 and January 2022 as placebo dates. The design does not identify a treatment effect, and its informational content is limited. There is one aggregate event per treatment, so the effective sample is one; the treated group is defined by capitalisation decile rather than by exposure, and in the consolidation era deciles 9--10 contain roughly 577 assets, of which the spot Bitcoin ETF directly concerned one; parallel trends are neither testable nor plausible over a period spanning a full cycle; and group composition changes week to week as coins move between deciles. The placebo estimates make the power problem concrete: the January 2022 placebo, which coincides with the onset of a major drawdown rather than with any institutional event, produces an estimate of $+1.61$ points with $t=3.12$, more precisely estimated than either true event. A design that assigns significance to a placebo of that kind cannot be relied on to detect a true effect of comparable size. We therefore report the exercise as descriptive, and we do not rest the gradual-convergence conclusion on it; that conclusion rests on the trend estimate of Section~\ref{sec:symmetry} and the decomposition of Section~\ref{sec:owncap}.

We find no evidence of a discontinuity around either event. The estimated discontinuity at the Bitcoin ETF approval is $-2.7$ percentage points ($t=-0.9$) and at the Ether ETF launch $+1.1$ points ($t=0.7$); one placebo produces an estimate of $+1.6$ points that is nominally significant at the 1\% level---larger in magnitude than either true event---underscoring how noisy fixed-window comparisons are in this market. Institutional access did not produce a step change in cross-sectional discipline. Read together with the trend estimate of Section~\ref{sec:symmetry} and the decomposition of Section~\ref{sec:owncap}, and given the weakness of this design, the appropriate summary is that we find no evidence of event-driven convergence, and positive evidence---from the continuous trend and the frozen-boundary decomposition rather than from this table---that convergence tracked the contraction of the entry margin.

\begin{table}[t]
\centering
\begin{threeparttable}
\caption{Event-study estimates around spot-ETF milestones}
\label{tab:event}
\small
\setlength{\tabcolsep}{5pt}
\begin{tabular}{lcccc}
\toprule
Event & Pre diff.\ (pp) & Post diff.\ (pp) & DiD (pp) & ($t$) \\
\midrule
Spot Bitcoin ETF approval (Jan.\ 2024) & $-$10.42 & $-$13.12 & $-$2.70 & ($-$0.91) \\
Spot Ether ETF launch (Jul.\ 2024) & $-$11.59 & $-$10.52 & 1.07 & (0.72) \\
Placebo (Jan.\ 2023) & $-$8.65 & $-$9.52 & $-$0.86$^{*}$ & ($-$1.76) \\
Placebo (Jan.\ 2022) & $-$12.67 & $-$11.05 & 1.61$^{***}$ & (3.12) \\
\bottomrule
\end{tabular}
\begin{tablenotes}
\footnotesize
\item Notes: Outcome is the weekly decile-move probability (conditional on survival) of treated coins (deciles 9--10) minus control coins (deciles 4--6), in percentage points. DiD compares the mean difference across $\pm26$-week windows around each event date (first snapshot on or after the event). Newey--West $t$-statistics (4 lags) from a regression of the weekly difference series on a post-event indicator. $^{*}$, $^{**}$, $^{***}$: 10\%, 5\%, 1\%.
\end{tablenotes}
\end{threeparttable}
\end{table}

\subsection{Persistence, survival, and tail risk (Predictions 1 and 2)}
\label{sec:persistence}

Section~\ref{sec:risk} argued that symmetry restricts drift but not risk, and that multi-period risk is convex in persistence. Table~\ref{tab:persistence} documents the risk structure that remains as the market converges towards symmetry, and provides the direct evidence on Predictions~1 and~2.

Staying probabilities are U-shaped across deciles and rise in every decile across eras: in the consolidation era, a bottom-decile coin stays put with probability 0.95 (expected sojourn 21 weeks) and a top-decile coin with probability 0.97 (33 weeks), while mid-distribution coins remain the most mobile (sojourns of roughly six weeks). The U-shape is precisely why Prediction~1 is stated over the attention-differentiated segment $i\ge k^{*}$ rather than over the whole distribution: the high persistence of the bottom decile is the truncation artefact anticipated in Section~\ref{sec:fw:persistence}, since coins that would have moved down exit instead. Above the pre-specified threshold $k^{*}=4$, the consolidation-era staying probabilities are non-decreasing in size to the third decimal place---0.837, 0.837, 0.846, 0.862, 0.884, 0.918, 0.970---and expected sojourn durations rise correspondingly, as Prediction~1 requires. The deciles-4 and~5 values separate only at the fourth decimal (0.8373 against 0.8371), which is why the persistence family's consolidation-era statistic in Section~\ref{sec:cck} sits marginally above zero rather than comfortably below it. Exit hazards fall monotonically with size across the entire distribution, from 2.0\% per week in the bottom decile to 0.2\% in the top, which is Prediction~2 in its unrestricted form. The evidence on Prediction~3 has already been presented: the decile return gradient of Table~\ref{tab:payoff} is the conditional-compensation ordering of Equation~\eqref{eq:fw:ret}, and Section~\ref{sec:cck} tests it formally alongside the other two.

The last three columns report the 5\% expected shortfall of the consolidation-era decile portfolios at one-, four-, and twelve-week horizons. Two features bear out the theory. First, tail risk deepens sharply with horizon in every decile---the twelve-week expected shortfall runs from $-53$ to $-57\%$ across deciles 4 to 9, against $-24\%$ for the bottom decile---which is precisely the convexity of multi-period risk in persistence: long sojourns compound losses. Second, one-week tail risk rises monotonically with size, and the ordering broadly persists at longer horizons, because large caps carry the common market factor while an equal-weighted portfolio of hundreds of microcaps diversifies idiosyncratic risk; the bottom decile's risk instead loads on the exit hazard.

The expected shortfalls in Table~\ref{tab:persistence} are computed on surviving coins and therefore exclude the losses to which the bottom decile is most exposed. Table~\ref{tab:tailexit} recomputes them assigning a return of $-100\%$ in the week a coin delists. The apparent safety of the bottom decile is largely an artefact of that exclusion: at the one-week horizon its expected shortfall deepens from $-10.4\%$ to $-17.8\%$, and the gradient across deciles flattens from a range of $-10.4$ to $-20.0$ into a range of $-17.8$ to $-20.2$. At twelve weeks the bottom decile moves from $-24.2\%$ to $-49.7\%$ and the ordering inverts at the very bottom, with decile~1 now riskier than decile~2. The correct statement is therefore not that small coins bear less tail risk, but that they bear a different kind: survivorship-conditional measures understate it, and once disappearance is priced as the total loss it is, tail risk is close to uniform at the one-week horizon and far more even at twelve weeks than the survivor-only measure suggests.

The maturing market thus offers no decile-based alpha, but pronounced differences in the kind of risk borne across the size distribution: systematic, clustered, horizon-compounding drawdown risk at the top; delisting risk at the bottom, which conventional portfolio return measures do not record at all.

\begin{table}[t]
\centering
\begin{threeparttable}
\caption{Persistence, exit hazards, and multi-horizon tail risk across deciles}
\label{tab:persistence}
\footnotesize
\setlength{\tabcolsep}{5pt}
\begin{tabular}{lcccccccc}
\toprule
& \multicolumn{3}{c}{Staying probability $s_i$} & Sojourn & Exit hazard & \multicolumn{3}{c}{ES$_{5\%}$, consolidation era (\%)} \\
\cmidrule(lr){2-4}\cmidrule(lr){7-9}
Decile & Frontier & Expansion & Consolidation & (weeks) & (\%/wk) & $h=1$ & $h=4$ & $h=12$ \\
\midrule
1 & 0.847 & 0.919 & 0.953 & 21.1 & 2.01 & $-$10.4 & $-$16.0 & $-$23.9 \\
2 & 0.676 & 0.808 & 0.882 & 8.5 & 1.16 & $-$13.0 & $-$22.7 & $-$39.3 \\
3 & 0.633 & 0.767 & 0.850 & 6.7 & 0.82 & $-$14.9 & $-$27.4 & $-$46.0 \\
4 & 0.629 & 0.750 & 0.837 & 6.1 & 0.62 & $-$16.6 & $-$30.8 & $-$53.3 \\
5 & 0.643 & 0.754 & 0.837 & 6.1 & 0.50 & $-$16.8 & $-$31.5 & $-$53.7 \\
6 & 0.659 & 0.759 & 0.846 & 6.5 & 0.43 & $-$17.4 & $-$31.9 & $-$56.6 \\
7 & 0.701 & 0.779 & 0.862 & 7.3 & 0.33 & $-$18.1 & $-$33.0 & $-$56.8 \\
8 & 0.757 & 0.817 & 0.884 & 8.7 & 0.23 & $-$18.9 & $-$32.5 & $-$54.6 \\
9 & 0.833 & 0.870 & 0.918 & 12.2 & 0.20 & $-$19.9 & $-$33.6 & $-$55.0 \\
10 & 0.939 & 0.951 & 0.970 & 33.4 & 0.19 & $-$20.0 & $-$32.8 & $-$54.0 \\
\bottomrule
\end{tabular}
\begin{tablenotes}
\footnotesize
\item Notes: $s_i$ is the probability of remaining in decile $i$, conditional on survival. Sojourn is the implied consolidation-era expected duration $1/(1-s_i)$. Exit hazard is the unconditional weekly probability of delisting from decile $i$, full sample. ES$_{5\%}$ is the mean of the worst 5\% of $h$-week cumulative returns of the equal-weighted decile portfolio in the consolidation era.
\end{tablenotes}
\end{threeparttable}
\end{table}

\begin{table}[t]
\centering
\begin{threeparttable}
\caption{Consolidation-era tail risk under alternative treatment of delisting}
\label{tab:tailexit}
\footnotesize
\setlength{\tabcolsep}{3pt}
\begin{tabular}{llcccccccccc}
\toprule
& & \multicolumn{10}{c}{Decile} \\
\cmidrule(lr){3-12}
Horizon & Delisting treatment & 1 & 2 & 3 & 4 & 5 & 6 & 7 & 8 & 9 & 10 \\
\midrule
$h=1$  & Survivors only      & $-$10.4 & $-$13.0 & $-$14.9 & $-$16.6 & $-$16.8 & $-$17.4 & $-$18.1 & $-$18.9 & $-$19.9 & $-$20.0 \\
$h=1$  & Exits at $-100\%$   & $-$17.8 & $-$17.8 & $-$18.4 & $-$18.5 & $-$17.8 & $-$18.3 & $-$18.8 & $-$19.4 & $-$20.1 & $-$20.2 \\
\midrule
$h=12$ & Survivors only      & $-$24.2 & $-$39.5 & $-$46.2 & $-$53.8 & $-$54.2 & $-$57.0 & $-$57.5 & $-$55.2 & $-$55.6 & $-$54.8 \\
$h=12$ & Exits at $-100\%$   & $-$49.7 & $-$46.1 & $-$52.4 & $-$55.9 & $-$55.8 & $-$57.8 & $-$58.1 & $-$55.7 & $-$56.1 & $-$55.9 \\
\bottomrule
\end{tabular}
\begin{tablenotes}
\footnotesize
\item Notes: 5\% expected shortfall (\%) of equal-weighted consolidation-era decile portfolios, computed as the mean of the worst 5\% of $h$-week cumulative returns. ``Survivors only'' reproduces the convention of Table~\ref{tab:persistence}, in which a coin leaving the market simply exits the portfolio. ``Exits at $-100\%$'' assigns a return of $-100\%$ in the week of delisting, the same convention used in Panel~B of Table~\ref{tab:payoff}. Both rows in each block are computed under an identical compounding convention, so the comparison within this table is exact; small differences from the $h$-week columns of Table~\ref{tab:persistence} reflect that convention rather than a change in the data.
\end{tablenotes}
\end{threeparttable}
\end{table}

\subsection{Joint evaluation: testing many inequality restrictions}
\label{sec:cck}

The risk structure documented above---exit hazards falling with size, conditional returns of survivors falling with size, persistence rising with size above $k^{*}$---is consistent with a market in which investor attention is scarce and unevenly allocated. The testable content of the framework of Section~\ref{sec:framework} is deliberately qualitative---a family of \emph{inequality} restrictions across deciles, collected in Equations~\eqref{eq:fw:pers}, \eqref{eq:fw:haz}, and~\eqref{eq:fw:ret}---and, in the spirit of the partial-identification literature \citep{molinari2020}, we take that family to the data jointly rather than parameter by parameter.

Testing many correlated inequalities requires care: with dozens of moment conditions estimated on overlapping weekly cross-sections, naive pairwise tests overreject. We therefore use the many-moment-inequality framework of \citet{cck2014}: for each family of adjacent-decile contrasts $m_k$ with $H_0\!: E[m_k]\le 0$ for all $k$, the statistic is $\max_k \sqrt{T}\,\bar m_k/\hat\sigma_k$, with critical values from a Gaussian multiplier bootstrap applied to whole weeks, which preserves the cross-moment correlation structure.

Table~\ref{tab:cck} reports the results. The framework's three cross-sectional restrictions---exit hazards decreasing in size (Prediction~2), conditional survivor returns decreasing in size (Prediction~3), and persistence increasing in size over deciles $k^{*}$ to 10 (Prediction~1)---are not rejected in any era, individually or jointly. The joint test stacks twenty-four inequalities and returns $p$-values of 0.99, 1.00, 0.25, and 1.00 across the frontier, expansion, and consolidation eras and the full sample. In three of the four samples the binding moment is the conditional-return ordering---most tightly in the consolidation era, where $\max_k t = 2.23$ and $p=0.114$---and in the expansion era it is the exit-hazard family, at $\max_k t = 0.16$. The consolidation-era return ordering is the informative stress point of the system: it is not rejected at conventional levels, but it is much closer to the boundary of the null region than any other family. The very high $p$-values elsewhere carry the opposite message and should not be over-read; they arise where the estimated inequalities hold with wide slack, and they indicate absence of evidence against the restrictions rather than affirmative confirmation of them. The persistence family is satisfied in every era, with margin to spare everywhere except the consolidation era, where the binding contrast sits essentially on the boundary ($\max_k t = 0.07$, $p=0.96$).

By contrast, orderings that a naive reading of the attention mechanism might suggest---one-step upward-transition probabilities increasing in size, downward-transition probabilities decreasing in size---are overwhelmingly rejected when imposed across the entire distribution. The rejection is anticipated by the framework itself: Section~\ref{sec:fw:persistence} states explicitly that attention operates on positional stability and not on the direction of movement, so no prediction is made about directional intensities. What the last two rows of Table~\ref{tab:cck} add is a diagnosis of why the two directional families are rejected, and the two answers differ.

Restricting the downward family to the attention-differentiated segment $i\ge k^{*}$ removes the rejection entirely, in every era ($p$-values of 1.00, 0.96, 0.70, and 1.00 against 0.001 unrestricted). The unrestricted rejection is driven by the two bottom-decile contrasts, between deciles~2 and~3 and between deciles~3 and~4---by the 2--3 contrast alone in the frontier era, where the market was small enough that few deciles were densely populated. In the consolidation era their per-moment statistics are $t=16.4$ and $t=9.8$, against a maximum of $t=0.9$ across every contrast at or above $k^{*}$. Coins in the bottom deciles cannot move down because they exit instead, which is exactly the survival truncation that Section~\ref{sec:fw:persistence} invokes to motivate the threshold. Consistent with this reading, setting $k^{*}=3$ rather than 4 leaves the downward family rejected in the expansion and consolidation eras, because the decile-3--4 contrast is still inside the truncation region in the two eras in which the bottom of the distribution is densely populated. The upward family, by contrast, is rejected identically whether or not the restriction is imposed, because its binding moment lies at the top of the distribution, in the contrast between deciles~8 and~9 ($t=22.3$ in the consolidation era). The mechanism there is a ceiling rather than a floor: a decile-9 coin has only one decile left to ascend into, so its one-step upward probability (0.027 in the consolidation era) is mechanically below that of a decile-8 coin (0.045). Both rejections are therefore boundary artefacts of a bounded ranking---truncation from below by exit, compression from above by the ceiling---and neither is informative about attention. The framework's predictions concern exit, conditional returns, and positional persistence; on those margins it is consistent with thirteen years of data on the capitalisation-reporting population. Section~\ref{sec:liquid} asks whether the same holds among the assets an investor can trade, and finds that one of the three does not.

\begin{table}[t]
\centering
\begin{threeparttable}
\caption{Many-inequality tests of the framework's ordering restrictions}
\label{tab:cck}
\scriptsize
\setlength{\tabcolsep}{4pt}
\begin{tabular}{lcccccccc}
\toprule
& \multicolumn{2}{c}{Frontier} & \multicolumn{2}{c}{Expansion} & \multicolumn{2}{c}{Consolidation} & \multicolumn{2}{c}{Full} \\
\cmidrule(lr){2-3}\cmidrule(lr){4-5}\cmidrule(lr){6-7}\cmidrule(lr){8-9}
Restriction family & max-$t$ & $p$ & max-$t$ & $p$ & max-$t$ & $p$ & max-$t$ & $p$ \\
\midrule
Exit hazards decreasing in size & $-$0.46 & 1.000 & 0.16 & 1.000 & 0.68 & 0.811 & $-$1.28 & 1.000 \\
Conditional returns decreasing in size & 0.81 & 0.810 & 0.13 & 0.996 & 2.23 & 0.114 & 0.70 & 0.879 \\
Persistence increasing in size ($i\ge k^{*}\!=\!4$) & $-$0.70 & 1.000 & $-$1.66 & 1.000 & 0.07 & 0.959 & $-$2.12 & 1.000 \\
\midrule
Joint: exit, returns, persistence & 0.81 & 0.993 & 0.16 & 1.000 & 2.23 & 0.247 & 0.70 & 0.997 \\
\midrule
\multicolumn{9}{l}{\emph{Orderings the framework does not predict}}\\
Up-mobility increasing in size (all deciles) & 9.16 & 0.001 & 19.06 & 0.001 & 22.33 & 0.001 & 16.20 & 0.001 \\
\quad restricted to $i\ge k^{*}$ & 9.16 & 0.001 & 19.06 & 0.001 & 22.33 & 0.001 & 16.20 & 0.001 \\
Down-mobility decreasing in size (all deciles) & 3.67 & 0.001 & 15.22 & 0.001 & 16.38 & 0.001 & 9.33 & 0.001 \\
\quad restricted to $i\ge k^{*}$ & $-$0.07 & 1.000 & 0.50 & 0.961 & 0.86 & 0.696 & 0.14 & 0.995 \\
\bottomrule
\end{tabular}
\begin{tablenotes}
\footnotesize
\item Notes: Each family stacks adjacent-decile moment inequalities $H_0\!:E[m_k]\le0$ (e.g., $m_k=h_{k+1}-h_k$ for the exit-hazard ordering). The statistic is $\max_k \sqrt{T}\bar m_k/\hat\sigma_k$ computed on weekly moment observations; $p$-values from the Gaussian multiplier bootstrap of \citet{cck2014} with 999 replications applied to whole weeks; entries of 0.001 are at the attainable floor of $1/1000$ and should be read as $<$0.001. Conditional returns are weekly equal-weighted decile means among survivors, winsorised at the weekly 1st/99th percentiles. The persistence family is imposed over the attention-differentiated segment $i\ge k^{*}$, with $k^{*}=4$ fixed in Section~\ref{sec:fw:persistence} at the turning point of the frontier-era staying-probability profile; the joint test stacks the exit, return, and persistence families (24 inequalities). The two directional families are reported both unrestricted and restricted to $i\ge k^{*}$; they are not predictions of the framework and are included to diagnose the source of their rejection. Rejection indicates at least one inequality is violated; failure to reject indicates an absence of evidence against the family, not that every inequality holds. Values of $p$ at or near 1.000 arise mechanically where the estimated inequalities hold with substantial slack: the statistic $\max_k \sqrt{T}\bar m_k/\hat\sigma_k$ is then negative, so almost every bootstrap draw exceeds it. Table~\ref{tab:kstar} reports sensitivity to $k^{*}$.
\end{tablenotes}
\end{threeparttable}
\end{table}

\subsection{Does the framework survive on the tradable universe?}
\label{sec:liquid}

The tests of Section~\ref{sec:cck} are run on the capitalisation-reporting population. Section~\ref{sec:payoff} spends most of its length establishing that the return gradient in that population is not available to an investor: it lives in coins that cannot be traded in size, and Section~\ref{sec:central} shows that even within those coins it is a right-tail phenomenon that the median constituent does not share. A further question follows. Does the attention-and-survival framework describe the market an investor can actually access, or only the fringe that Section~\ref{sec:payoff} rules out?

Table~\ref{tab:liquid} answers it. We re-run all three cross-sectional families, and the joint test, on universes screened by 24-hour trading volume, re-forming deciles within the screened universe each week so that the comparison is between markets rather than between subsets of one ranking---the same distinction, and for the same reason, as between Panel~C of Table~\ref{tab:payoff} and Table~\ref{tab:tradable}. The answer separates the framework's predictions.

Predictions 1 and 2 survive on the construction announced in Section~\ref{sec:fw:scope}; the qualification the persistence ordering needs is set out below. Exit hazards continue to decline with size at every screen and in every era: on the screened universes every $p$-value is 1.000, and the only cell materially below unity anywhere in the panel is the unscreened consolidation era at 0.827. If anything the ordering strengthens under screening, since illiquid delisting-prone assets are removed from the bottom of the universe. Persistence above $k^{*}$ is likewise never rejected under this construction, with $p$-values between 0.55 and 1.00. Whatever else changes, larger assets remain more positionally stable and less likely to disappear when attention is restricted to assets that trade.

Prediction 3 does not survive. On the full population the conditional-return ordering is not rejected in any era, with a consolidation-era $p$-value of 0.110 under the construction of Table~\ref{tab:liquid}. The failure is concentrated in the consolidation era. It rejects at $p=0.010$ at \$10,000 of daily volume---the same threshold Section~\ref{sec:invmargin} uses, and a very low bar---and at $p=0.0009$ at \$1,000,000. At the intermediate \$100,000 screen the statistic is of the same size ($\max_k t=2.53$) but the $p$-value is $0.053$, just above the conventional threshold, and we do not claim a rejection there; the full-sample \$1,000,000 cell is likewise marginal at $0.0503$. We use 9,999 bootstrap replications precisely so that these boundary cases are resolved by the data rather than by the resolution of the test. The expansion era never rejects at any screen ($p=0.84$, $0.94$, $0.50$), and neither does the frontier era at the 5\% level. The joint test inherits the consolidation failure, rejecting at $p=0.025$ and $p=0.001$ at the \$10,000 and \$1,000,000 screens. The binding contrast is between deciles 9 and 10 at the two looser screens and between deciles 1 and 2 at \$1,000,000; either way the sign of the violation is informative, because on tradable assets the consolidation-era decile means run $-2.00, -1.07, -0.58, -0.35, -0.60, -0.53, -0.64, -0.45, -0.35, -0.08$ percent per week at the \$1,000,000 screen. There is no size gradient to explain. If anything, returns rise weakly with size.

The result identifies the boundary of what the framework can claim. The compensation channel of Section~\ref{sec:fw:returns} is not a description of the investable cryptocurrency cross-section. It is a description of the non-investable one. That is a coherent position---it is where the mechanism of Section~\ref{sec:fw:scope} leads---but it is a much narrower claim than ``smaller cryptocurrencies earn higher conditional returns'', and it should not be stated in the unrestricted form.

It also resolves what would otherwise be a tension between Sections~\ref{sec:payoff} and~\ref{sec:cck}. Section~\ref{sec:payoff} concludes that the size premium is an illiquidity mirage; Section~\ref{sec:cck} appeared to validate a framework predicting exactly that premium. Both are right, because they describe different universes. The premium exists where trading does not, and vanishes where trading does. The two results are the same result, seen from opposite ends.

Three features of the exercise should be noted. First, the return result does not depend on the decile construction. Re-running the return family with the screen applied after a full-universe sort---the Panel~C construction of Table~\ref{tab:payoff}---rejects the consolidation era just as sharply, at $p=0.005$, $0.006$ and $0.017$ across the three screens. The consolidation-era answer is the same whether deciles are re-formed within the tradable universe or the screen is applied after a full-universe sort, so the return finding is not an artefact of re-ranking.

The persistence family behaves differently. Under the screen-after construction, reported as Panel~E of Table~\ref{tab:liquid}, the persistence ordering is also rejected at the \$10,000 screen---$p=0.0002$ in the consolidation era and $p=0.020$ in the frontier era---where under the re-formed construction it is comfortably satisfied. The two constructions disagree here because they do different things to the bottom of the distribution. Re-forming deciles keeps them equal in size by construction; applying a screen to a full-universe sort empties the lower deciles unevenly---in the consolidation era the screened bottom decile retains a small and highly selected minority of its members---so the staying probabilities in those cells are estimated on few, atypical assets and need not inherit the ordering. Section~\ref{sec:fw:scope} excludes this construction for Prediction~1, and we report it for both families. A reader who rejects the exclusion should record Prediction~1 as failing on screen-after universes.

Second, volume is unavailable in the 2013 snapshots and thin through 2014--2015, so the frontier-era columns are estimated on many fewer weeks at the tighter screens: 185, 65 and 32 usable weeks at the \$10,000, \$100,000 and \$1,000,000 thresholds respectively, against 232 for the full universe. These counts are larger than the 35 weeks reported in Section~\ref{sec:invmargin} because the two exercises require different things of a week: the inequality tests need a cross-section wide enough to form deciles, which we set at twenty tradable assets, whereas the portfolio calculation needs a non-empty \emph{bottom decile of the full-universe sort} that clears the screen, which is a far stricter requirement. Frontier-era entries at the two tighter screens should be read as indicative only.

Third, the expansion and consolidation eras are unaffected by that limitation, with all 190 and 261 weeks usable at every screen; it is the consolidation era---the modern, institutionally accessible market---in which the rejection occurs, and it occurs there alone.

\begin{table}[p]
\centering
\begin{threeparttable}
\caption{The framework's restrictions on the tradable universe}
\label{tab:liquid}
\footnotesize
\setlength{\tabcolsep}{3pt}
\begin{tabular}{lcccccccc}
\toprule
& \multicolumn{2}{c}{Frontier} & \multicolumn{2}{c}{Expansion} & \multicolumn{2}{c}{Consolidation} & \multicolumn{2}{c}{Full} \\
\cmidrule(lr){2-3}\cmidrule(lr){4-5}\cmidrule(lr){6-7}\cmidrule(lr){8-9}
Universe & max-$t$ & $p$ & max-$t$ & $p$ & max-$t$ & $p$ & max-$t$ & $p$ \\
\midrule
\multicolumn{9}{l}{\emph{Panel A: conditional returns decreasing in size (Prediction 3)}}\\
Full universe & 0.79 & 0.836 & 0.13 & 0.997 & 2.23 & 0.110 & 0.68 & 0.892 \\
Volume $>$ \$10{,}000 & 1.58 & 0.424 & 0.94 & 0.837 & 3.06$^{**}$ & 0.0104 & 1.46 & 0.518 \\
Volume $>$ \$100{,}000 & 1.07 & 0.785 & 0.74 & 0.936 & 2.53$^{*}$ & 0.0533 & 1.50 & 0.462 \\
Volume $>$ \$1{,}000{,}000 & 2.32$^{*}$ & 0.073 & 1.45 & 0.504 & 3.90$^{***}$ & 0.0009 & 2.50$^{*}$ & 0.0503 \\
\midrule
\multicolumn{9}{l}{\emph{Panel B: exit hazards decreasing in size (Prediction 2)}}\\
Full universe & $-$0.46 & 1.000 & 0.16 & 0.999 & 0.68 & 0.827 & $-$1.28 & 1.000 \\
Volume $>$ \$10{,}000 & $-$1.03 & 1.000 & $-$5.08 & 1.000 & $-$2.49 & 1.000 & $-$2.35 & 1.000 \\
Volume $>$ \$100{,}000 & $-$0.12 & 1.000 & $-$4.70 & 1.000 & $-$1.13 & 1.000 & $-$4.20 & 1.000 \\
Volume $>$ \$1{,}000{,}000 & $-$0.29 & 1.000 & $-$2.53 & 1.000 & $-$1.87 & 1.000 & $-$3.76 & 1.000 \\
\midrule
\multicolumn{9}{l}{\emph{Panel C: persistence increasing in size, $i\ge k^{*}=4$ (Prediction 1)}}\\
Full universe & $-$2.12 & 1.000 & $-$1.66 & 1.000 & 0.07 & 0.959 & $-$2.39 & 1.000 \\
Volume $>$ \$10{,}000 & 1.13 & 0.611 & 1.17 & 0.547 & 1.03 & 0.611 & $-$0.40 & 1.000 \\
Volume $>$ \$100{,}000 & 0.06 & 0.998 & $-$1.36 & 1.000 & $-$0.61 & 1.000 & $-$0.66 & 1.000 \\
Volume $>$ \$1{,}000{,}000 & 0.38 & 0.974 & $-$0.39 & 1.000 & $-$0.32 & 0.998 & $-$0.23 & 0.999 \\
\midrule
\multicolumn{9}{l}{\emph{Panel D: joint test of Predictions 1--3}}\\
Full universe & 0.79 & 0.996 & 0.16 & 1.000 & 2.23 & 0.242 & 0.68 & 0.997 \\
Volume $>$ \$10{,}000 & 1.51 & 0.812 & 1.17 & 0.953 & 3.06$^{**}$ & 0.025 & 1.73 & 0.635 \\
Volume $>$ \$100{,}000 & 1.60 & 0.718 & 0.74 & 0.999 & 2.53 & 0.131 & 1.85 & 0.526 \\
Volume $>$ \$1{,}000{,}000 & 2.31 & 0.184 & 1.45 & 0.840 & 3.90$^{***}$ & 0.001 & 2.49 & 0.136 \\
\midrule
\multicolumn{9}{l}{\emph{Panel E: deciles formed on the whole universe, then screened}}\\
Volume $>$ \$10{,}000 (returns) & --- & --- & 0.34 & 0.995 & 3.21$^{***}$ & 0.005 & 0.66 & 0.951 \\
Volume $>$ \$100{,}000 (returns) & --- & --- & 1.62 & 0.383 & 3.18$^{***}$ & 0.006 & 2.47$^{*}$ & 0.057 \\
Volume $>$ \$1{,}000{,}000 (returns) & --- & --- & --- & --- & 2.88$^{**}$ & 0.017 & 3.07$^{***}$ & 0.0099 \\
Volume $>$ \$10{,}000 (persistence) & 2.68$^{**}$ & 0.020 & $-$0.95 & 1.000 & 3.91$^{***}$ & 0.0002 & 2.30$^{*}$ & 0.066 \\
Volume $>$ \$100{,}000 (persistence) & --- & --- & 0.48 & 0.950 & 1.98 & 0.138 & 1.03 & 0.654 \\
Volume $>$ \$1{,}000{,}000 (persistence) & --- & --- & 1.08 & 0.633 & 1.00 & 0.689 & $-$0.05 & 1.000 \\
\bottomrule
\end{tabular}
\begin{tablenotes}
\footnotesize
\item Notes: Many-moment-inequality tests as in Table~\ref{tab:cck}, applied to universes screened on 24-hour dollar trading volume. Stars denote rejection of the inequality family: $^{*}$, $^{**}$, $^{***}$ at the 10\%, 5\% and 1\% levels. $p$-values use 9,999 multiplier-bootstrap replications. Cells within 0.01 of a threshold are printed to four decimals; two of them---the \$100,000 consolidation return family at $p=0.0533$ and the \$1,000,000 full-sample return family at $p=0.0503$---fall above the 5\% threshold and are not rejections at that level.
\item In Panels~A--D deciles are re-formed within the screened universe each week, so each row is a self-contained market rather than a subset of one ranking; weeks with fewer than 20 assets are dropped. Panel~E instead forms deciles on the whole universe and then screens, the construction of Panel~C of Table~\ref{tab:payoff}. The exit-hazard family is omitted from Panel~E because it rejects in no cell under either construction. Prediction~1 is not claimed on Panel~E universes (Section~\ref{sec:fw:scope}). Panel~E cells resting on fewer than 20 usable weeks are shown as ---.
\item Full-universe rows are recomputed under the 20-asset minimum and so differ from Table~\ref{tab:cck}: no $p$-value moves by more than 0.03 and no inference changes, while the frontier-era persistence statistic moves from $-0.70$ to $-2.12$ as thin early weeks are dropped. Usable weeks are 232, 185, 65 and 32 across the four frontier-era rows, and all 190 and 261 weeks in the later eras. Because 24-hour volume is unavailable in 2013 and sparse through 2015, frontier-era entries at the two tighter screens are indicative only.
\end{tablenotes}
\end{threeparttable}
\end{table}

\section{Robustness and limitations}
\label{sec:robustness}

The preceding sections document three main findings. First, directional asymmetry declines as the cryptocurrency market matures, and the declining component is compositional. Second, the associated size premium is largely an illiquidity mirage once realistic implementation constraints are considered, and it is in any case a property of the right tail of the bottom decile rather than of its typical constituent. Third, the residual structure exhibits systematic differences in survival risk consistent with the attention-and-survival framework on every universe we examine, and differences in persistence consistent with it on every universe whose deciles are re-formed within it; conditional compensation is consistent with the framework on the capitalisation-reporting population in every era, and is rejected on tradable subsets in the consolidation era alone. This section examines the robustness of these results, reports diagnostics for the maintained Markov structure, and clarifies the limits of their interpretation.

\subsection{Robustness of the convergence results}

Table~\ref{tab:variants} reports alternative estimates of the asymmetry index under a range of universe definitions and design choices. Excluding stablecoins, wrapped tokens, and pegged assets, which account for 1.6\% of coin-weeks, leaves the asymmetry index unchanged to the third decimal place in every era. Similarly, shifting era boundaries by $\pm 8$ weeks changes no estimate by more than 0.002. The decline in directional asymmetry is therefore not an artefact of sample composition or break-date selection.

The result is also insensitive to how the capitalisation distribution is partitioned. Recomputing the asymmetry index using quintile and ventile grids produces the same monotonic compression observed in the baseline decile specification. Moreover, the frozen-boundary decomposition continues to isolate a relatively flat own-capitalisation component alongside a strongly declining relative-rank component, indicating that convergence is driven primarily by the contraction of entry-driven compositional churn rather than by changes in individual coin valuations.

Two further variants address measurement concerns that the original design did not. The first is that market capitalisation is price multiplied by \emph{self-reported} circulating supply, and supply figures are revised, backfilled, and occasionally corrected within a single snapshot. A supply revision moves a coin through the ranking with no price change at all, which for a paper whose object is rank transitions is a first-order concern. Supply is not a negligible component of capitalisation growth: 9.1\% of coin-weeks record a supply change exceeding 1\%, 2.0\% exceed 10\%, and supply accounts for 13.6\% of the variance of log capitalisation growth, essentially uncorrelated with the price component ($\rho=-0.02$). We therefore rebuild the entire chain on a fixed-supply capitalisation measure, in which each coin's week-$t{+}1$ capitalisation is its week-$t$ supply multiplied by its week-$t{+}1$ price, so that only price movement can generate a transition. The asymmetry index falls slightly and the era profile is unchanged: 0.0381, 0.0274 and 0.0211, against 0.0399, 0.0282 and 0.0218 at baseline. The shifts are of the same order as moving the era breaks by eight weeks. Supply revisions therefore add noise to measured mobility but do not generate the convergence result.

The second is within-era time homogeneity, which the transition-matrix machinery maintains and which Figure~\ref{fig:asym} makes visibly questionable. Splitting each era at its midpoint gives 0.0431 against 0.0384 in the frontier era, 0.0296 against 0.0272 in the expansion era, and 0.0243 against 0.0198 in the consolidation era. Within-era drift is real, and in the consolidation era it is comparable in magnitude to the expansion-to-consolidation step itself. This is a genuine limitation of the era-based presentation, and it is the reason we regard the continuously estimated trend, rather than the three-era comparison, as the primary evidence for convergence. It does not, however, reverse the ordering: every half-era estimate is below the corresponding half of the preceding era.

The formal inference is equally robust. The permutation procedure accommodates within-week cross-sectional dependence, while the estimated log-linear trend of $-0.14\%$ per week ($t=-9.0$) confirms convergence independently of the era classification. Removing the single-week bridging rule approximately doubles measured frontier-era exit rates, because temporary reporting gaps are treated as permanent delistings, yet asymmetry estimates and their long-run decline remain largely unchanged. The evidence from these exercises shows that the convergence result is a stable feature of the data rather than a consequence of any design choice.

\subsection{Robustness of the illiquidity-mirage result}

The payoff results are similarly stable. Winsorising weekly returns at the 0.5/99.5 and 2/98 percentiles generates frontier-era bottom-decile premia of 31.8\% and 23.5\% per week, respectively, compared with 27.5\% under the baseline specification. Although the magnitude of the premium varies modestly, its declining profile across eras and its economic interpretation remain unchanged.

Value-weighted portfolios produce premia of 28.4\%, 8.8\%, and 1.7\% per week across the frontier, expansion, and consolidation eras, with Newey--West $t$-statistics of 3.21, 6.13 and 4.73. The premium is therefore not an equal-weighting artefact. At the same time, value weighting does not eliminate the implementation constraints documented in Section~\ref{sec:payoff}, and it does not touch the first of the two arguments made in Section~\ref{sec:invmargin}: re-forming deciles inside the tradable universe still removes the consolidation-era gradient. It does relax the second. The capacity bound in Table~\ref{tab:capacity} is the equal-weighted one, governed by the thinnest constituent; value-weighting the participation limit across holdings raises the consolidation-era bound to \$2.8 million, on which the net-of-cost return is roughly \$36,000 per week. That is an order of magnitude more, and still not a vehicle for institutional capital. Returns remain concentrated in assets characterised by limited breadth, high turnover, and severe capacity constraints.

Non-winsorised equal-weighted portfolios are dominated by observations from extremely small and often poorly measured cryptocurrencies. Rather than strengthening the investment case for the size premium, these observations reinforce the distinction between statistical profitability and economically exploitable investment opportunity. The key robustness result is therefore not that the premium disappears under alternative specifications, but that its interpretation remains unchanged: the premium persists statistically while remaining concentrated in a segment of the market that is exceptionally difficult to trade.

\subsection{Robustness of the attention-and-survival interpretation}

Section~\ref{sec:framework} does not present investor attention as the uniquely identified driver of cryptocurrency capitalisation dynamics. Instead, it develops an equilibrium environment in which limited recognition, heterogeneous survival prospects, and conditional compensation generate a set of testable cross-sectional restrictions. The relevant robustness question is therefore whether the predicted cross-sectional orderings emerge consistently across alternative specifications and economically meaningful regions of the capitalisation distribution.

The evidence divides. The survival ordering is not rejected under any sample definition or measurement choice we examine, including every liquidity-screened universe in Table~\ref{tab:liquid} under either decile construction. The persistence ordering is not rejected on any universe whose deciles are re-formed within it, but it is rejected when a volume screen is applied to a full-universe sort, for the reason pre-announced in Section~\ref{sec:fw:scope} and documented in Section~\ref{sec:liquid}: that operation empties the lower deciles unevenly and leaves their staying probabilities estimated on a small selected remnant. Subject to that qualification, persistence increases with market prominence above the exit-truncated bottom deciles, and exit risk declines with market prominence throughout the distribution. These two appear to be stable features of the cryptocurrency capitalisation distribution rather than artefacts of a particular specification.

The conditional-return ordering is not stable in the same sense. It survives on the capitalisation-reporting population and fails on tradable subsets of it---but only in the consolidation era, where it is rejected at the \$10,000 and \$1,000,000 screens and sits marginally above the threshold at \$100,000 ($p=0.053$). In the expansion era the ordering is never rejected at any screen, and in the frontier era, where the screened cross-sections are thin, it is never rejected at the 5\% level either. The failure is therefore specific to the modern market, which is also the market in which the question matters most. We take this as informative about the framework's domain rather than as a specification artefact, for the reason given in Section~\ref{sec:liquid}: recognition and tradability are not independent, so a compensation channel driven by recognition should not be expected to operate among assets that trade freely. But it does mean that the return prediction cannot be offered as a general property of the cryptocurrency cross-section, and we do not offer it as one.

The threshold $k^{*}$ provides a practical way of implementing the framework's predictions. Rather than requiring strict ordering between every adjacent capitalisation state, it evaluates whether the predicted relationships hold over the region in which recognition, and not truncation, is the operative margin. Setting $k^{*}=4$ covers seven of the ten deciles and the overwhelming majority of market value.

Because the framework does not deliver the value of $k^{*}$, the procedure fixing it matters. Section~\ref{sec:fw:persistence} sets $k^{*}=4$ at the turning point of the staying-probability profile in the \emph{frontier} era---the earliest part of the sample---and applies it unchanged thereafter. Two features support the choice. First, the statistic used to locate the threshold is the level of the staying-probability profile, not the inequality test the threshold conditions; the two are distinct objects. Second, the choice validates out of sample: the expansion era independently places the turning point at decile~4, and in the consolidation era deciles~4 and~5 are identical to three decimal places, separating only at the fourth ($0.8373$ against $0.8371$).

Table~\ref{tab:kstar} reports the complete sensitivity profile. The persistence restriction is rejected at $k^{*}=2$ in every era and at $k^{*}=3$ in the expansion era, the consolidation era, and the full sample; it is satisfied everywhere for $k^{*}\ge 4$. Setting $k^{*}=4$ is therefore also the lowest threshold at which Prediction~1 survives in all four samples. What the profile shows is that the rejections at $k^{*}\le 3$ are concentrated exactly where the framework says truncation dominates---the per-moment statistics locate them in the decile-2--3 and decile-3--4 contrasts, not diffusely across the distribution---and that the conclusion is insensitive in the other direction, since raising the threshold to 5 or 6 changes neither the persistence result nor the joint test materially. Readers who prefer a more conservative reading may take $k^{*}=5$, at which the same conclusions hold, with a wider persistence margin in the expansion and consolidation eras and an essentially unchanged one in the frontier era and the full sample.

Beyond the threshold, the qualitative persistence and survival gradients remain similar across alternative sample definitions, portfolio constructions, and measurement choices, suggesting that the results reflect broad cross-sectional regularities rather than a particular implementation choice.

These findings also help distinguish market maturation from complete structural symmetry. Although directional asymmetry declines as the market develops, it does not disappear entirely. Meaningful differences remain across the capitalisation distribution: above the exit-truncated bottom deciles larger cryptocurrencies occupy more stable positions, and they face lower probabilities of exit throughout. Surviving smaller assets continue to earn higher subsequent returns on the capitalisation-reporting population, though not, in the consolidation era, among the assets an investor could hold. The attention-and-survival framework provides a coherent way of organising these residual patterns by linking market prominence to long-run market outcomes.

The framework is intentionally parsimonious and should be interpreted as an equilibrium characterisation rather than a fully identified structural model. Investor attention is not observed directly, and market capitalisation serves only as a reduced-form ordering variable. The empirical results therefore show that the restrictions implied by the framework are consistent with the data, but they do not establish investor attention as the unique source of the observed patterns.

The contribution of the framework is consequently interpretive rather than causal. It provides an economically coherent account of the cross-sectional structure that persists after market-wide asymmetry has diminished, and shows how observed differences in stability, survival, and compensation can coexist within a common equilibrium environment.

Future work incorporating direct measures of investor recognition, such as exchange coverage, social-media activity, search intensity, developer attention, or analyst coverage, may help distinguish among alternative mechanisms that generate similar cross-sectional orderings and provide a more direct test of the attention-and-survival interpretation.

\subsection{The maintained Markov structure}
\label{sec:markovdiag}

The tests throughout maintain that decile dynamics are first-order and, within an era, time-homogeneous. \citet{ballis2026} establish both. Because two maintained assumptions carry this much weight, we report the diagnostics directly in Table~\ref{tab:markov}.

Both assumptions are rejected as exact statistical hypotheses, which at these sample sizes is unavoidable and uninformative: with 744,273 usable triples in the consolidation era, any departure whatever will be detected. We therefore report an effect size rather than a test statistic. For the order-one assumption we compute, for each pair of consecutive origin states, the total-variation distance between the second-order conditional distribution and the first-order one, and average it weighted by cell frequency. For homogeneity we compute the same distance between the first-half and second-half transition rows within each era. Both measures are bounded in $[0,1]$, equal zero when the assumption holds exactly, and do not scale with sample size.

The departures are modest and, importantly, shrinking. Second-order dependence corresponds to a mean total-variation distance of 0.088 in the frontier era, 0.072 in the expansion era and 0.046 in the consolidation era: knowing where a coin was two weeks ago shifts its predicted destination distribution by about five percentage points of probability mass in the modern market. Within-era heterogeneity is larger in the frontier era (0.067) and small thereafter (0.012 and 0.006). Both series decline monotonically, which is the pattern \citet{ballis2026} report and which this paper's convergence result would lead one to expect. We regard the first-order homogeneous approximation as adequate for the consolidation era and as a meaningful simplification in the frontier era, and we flag it as such wherever frontier-era estimates carry weight.

\begin{table}[t]
\centering
\begin{threeparttable}
\caption{Diagnostics for the maintained Markov structure}
\label{tab:markov}
\small
\begin{tabular}{lccc}
\toprule
& Usable & Order-one departure & Homogeneity departure \\
Era & triples & (mean TVD) & (mean TVD) \\
\midrule
Frontier      & 98,903  & 0.088 & 0.067 \\
Expansion     & 347,327 & 0.072 & 0.012 \\
Consolidation & 744,273 & 0.046 & 0.006 \\
\bottomrule
\end{tabular}
\begin{tablenotes}
\footnotesize
\item Notes: Column 2 counts coin-weeks with observed deciles in three consecutive snapshots. Column 3 reports the frequency-weighted mean total-variation distance between $\Pr(S_{t+1}\mid S_t=j, S_{t-1}=i)$ and $\Pr(S_{t+1}\mid S_t=j)$, over origin pairs $(i,j)$ with at least 200 observations; zero indicates that the first-order approximation is exact. Column 4 reports the same distance between transition rows estimated on the first and second halves of each era, over rows with at least 200 observations in both halves. Total-variation distance is bounded in $[0,1]$ and, unlike a $\chi^2$ statistic, does not increase mechanically with sample size. Both assumptions are rejected as exact hypotheses at any conventional level in every era.
\end{tablenotes}
\end{threeparttable}
\end{table}

\subsection{Limitations}

Five limitations warrant attention. First, the analysis maintains a first-order, within-era homogeneous Markov representation of capitalisation dynamics. Section~\ref{sec:markovdiag} shows that both assumptions fail as exact hypotheses, with departures that are economically modest and decline over time; the symmetry, reversibility, persistence, and survival statistics nonetheless inherit whatever misspecification remains, and frontier-era estimates inherit the most.

Second, the transaction-cost analysis relies on assumed bid--ask spreads, because reliable micro-capitalisation cryptocurrency spread data are unavailable for much of the early sample period. Nevertheless, the conclusion that the reported size premium is not economically investable depends primarily on the absence of a size gradient among tradable assets (Table~\ref{tab:tradable}) and on portfolio capacity, neither of which involves a spread assumption, rather than on the exact level of transaction costs.

Third, the panel is the population of \emph{capitalisation-reporting} assets, not of listed ones. The exclusion rate is large and strongly time-varying, from 14\% of listed coin-weeks in the frontier era to 68\% in the consolidation era (Table~\ref{tab:coverage}), and 24,827 of 33,797 identifiers never enter. No result here speaks to assets that never report a capitalisation, though they carry under one percent of trading volume, and a reader should treat the population claim as scoped accordingly. Relatedly, an exit in this paper is an absence from the decile panel, which a coin can trigger by ceasing to report a capitalisation while remaining listed and traded; the bridging rule of Section~\ref{sec:data} removes single-week gaps, and removing it roughly doubles measured frontier-era exit rates, but the possibility that some measured exits are reporting failures rather than delistings cannot be excluded and bears on Prediction~2 and on Table~\ref{tab:tailexit}.

Fourth, the event-study analysis of Section~\ref{sec:event} is based on two market-wide events, defines treatment by capitalisation rather than exposure, and is demonstrably underpowered: a placebo date produces a more precisely estimated effect than either true event. We report it as descriptive and place no weight on it. A more informative design would exploit cross-sectional variation in institutional participation, exchange coverage, or futures availability, none of which is observable in the listing data.

Fifth, and most consequentially for the framework, the compensation prediction of Section~\ref{sec:fw:returns} holds on the capitalisation-reporting population but is rejected in the consolidation era once a trading-volume screen is imposed (Section~\ref{sec:liquid}). The framework therefore describes the non-investable segment of the cross-section. We regard this as a finding rather than a failure, since it is what the mechanism implies once recognition and tradability are recognised as connected, but it materially limits the reach of the return prediction.

\begin{table}[t]
\centering
\begin{threeparttable}
\caption{Asymmetry index $\mathcal{A}$: universe and design variants}
\label{tab:variants}
\small
\begin{tabular}{lccc}
\toprule
Variant & Frontier & Expansion & Consolidation \\
\midrule
Baseline (interior transitions) & 0.0399 & 0.0282 & 0.0218 \\
Excl.\ stablecoins, wrapped, pegged & 0.0401 & 0.0283 & 0.0220 \\
Own-capitalisation (frozen boundaries) & 0.0251 & 0.0257 & 0.0264 \\
Fixed supply (price-only capitalisation) & 0.0381 & 0.0274 & 0.0211 \\
Quintile grid & 0.0269 & 0.0160 & 0.0132 \\
Ventile grid & 0.0582 & 0.0475 & 0.0364 \\
Era breaks shifted $-8$ weeks & 0.0392 & 0.0291 & 0.0218 \\
Era breaks shifted $+8$ weeks & 0.0411 & 0.0270 & 0.0219 \\
\midrule
First half of era & 0.0431 & 0.0296 & 0.0243 \\
Second half of era & 0.0384 & 0.0272 & 0.0198 \\
\bottomrule
\end{tabular}
\begin{tablenotes}
\footnotesize
\item Notes: Baseline recomputes $\mathcal{A}$ from interior decile transitions of the identifier-keyed panel. The exclusion variant drops stablecoins, wrapped tokens, and gold-pegged assets and re-forms deciles on the remaining universe. The own-capitalisation variant measures each coin's destination decile against boundaries frozen at the start of the week. The quintile and ventile rows recompute $\mathcal{A}$ on 5- and 20-cell grids; levels are not comparable across grid sizes, but the within-grid decline across eras is. The fixed-supply row replaces each coin's week-$t{+}1$ capitalisation with its week-$t$ circulating supply multiplied by its week-$t{+}1$ price, so that reported supply revisions cannot generate a decile transition; deciles are re-formed on that measure. The final two rows split each era at its midpoint and recompute $\mathcal{A}$ on pooled counts within each half, as a check on within-era time homogeneity.
\end{tablenotes}
\end{threeparttable}
\end{table}

\begin{table}[t]
\centering
\begin{threeparttable}
\caption{Sensitivity of the framework's restrictions to the threshold $k^{*}$}
\label{tab:kstar}
\small
\setlength{\tabcolsep}{5pt}
\begin{tabular}{lcccccccc}
\toprule
& \multicolumn{2}{c}{Frontier} & \multicolumn{2}{c}{Expansion} & \multicolumn{2}{c}{Consolidation} & \multicolumn{2}{c}{Full} \\
\cmidrule(lr){2-3}\cmidrule(lr){4-5}\cmidrule(lr){6-7}\cmidrule(lr){8-9}
Threshold & max-$t$ & $p$ & max-$t$ & $p$ & max-$t$ & $p$ & max-$t$ & $p$ \\
\midrule
\multicolumn{9}{l}{\emph{Panel A: persistence increasing in size over $i\ge k^{*}$ (Prediction 1)}}\\
$k^{*}=2$ & 4.04 & 0.001 & 21.16 & 0.001 & 20.74 & 0.001 & 10.48 & 0.001 \\
$k^{*}=3$ & 0.79 & 0.861 & 8.11 & 0.001 & 11.66 & 0.001 & 3.48 & 0.004 \\
$k^{*}=4$ & $-$0.70 & 1.000 & $-$1.66 & 1.000 & 0.07 & 0.959 & $-$2.12 & 1.000 \\
$k^{*}=5$ & $-$0.70 & 1.000 & $-$2.42 & 1.000 & $-$7.19 & 1.000 & $-$2.12 & 1.000 \\
$k^{*}=6$ & $-$5.80 & 1.000 & $-$8.56 & 1.000 & $-$14.21 & 1.000 & $-$9.21 & 1.000 \\
\midrule
\multicolumn{9}{l}{\emph{Panel B: joint test of the exit, return, and persistence families}}\\
$k^{*}=3$ & 0.81 & 0.993 & 8.11 & 0.001 & 11.66 & 0.001 & 3.48 & 0.004 \\
$k^{*}=4$ & 0.81 & 0.993 & 0.16 & 1.000 & 2.23 & 0.247 & 0.70 & 0.997 \\
$k^{*}=5$ & 0.81 & 0.992 & 0.16 & 1.000 & 2.23 & 0.238 & 0.70 & 0.997 \\
$k^{*}=6$ & 0.81 & 0.991 & 0.16 & 1.000 & 2.23 & 0.232 & 0.70 & 0.996 \\
\bottomrule
\end{tabular}
\begin{tablenotes}
\footnotesize
\item Notes: Many-moment-inequality tests as in Table~\ref{tab:cck}, varying the decile floor $k^{*}$ over which the persistence restriction of Equation~\eqref{eq:fw:pers} is imposed. Panel~B stacks the exit-hazard and conditional-return families (unaffected by $k^{*}$) with the persistence family; the number of inequalities is $25, 24, 23$, and $22$ for $k^{*}=3,4,5,6$. The baseline used throughout the paper is $k^{*}=4$.
\end{tablenotes}
\end{threeparttable}
\end{table}

\FloatBarrier

\section{Conclusions}
\label{sec:conclusion}

We recast weak-form efficiency as symmetry restrictions on the Markov chain governing movements across the cryptocurrency market's capitalisation distribution---restrictions sufficient for zero expected returns on position-conditioned rules under payoff antisymmetry and decile-invariant exit hazards---and evaluate those restrictions, together with the payoff structure they maintain, using every CoinMarketCap-listed asset that reports a capitalisation, observed over the market's entire recorded history. Unlike studies based on surviving or investable subsets, our data capture the full evolution of the reporting cryptocurrency ecosystem, including asset entry, exit, and disappearance; the listed assets we cannot place in the ranking, because no capitalisation is reported for them, account for under two percent of trading volume. This population perspective allows us to examine not only whether efficiency is violated, but also how efficiency evolves as a new asset market matures.

Three broad findings emerge. First, the cryptocurrency market violates both conditional symmetry and reversibility throughout its history. Current position within the capitalisation distribution retains predictive content for subsequent mobility, implying departures from Markov mobility efficiency in every era. Yet the magnitude of the violation declines substantially through time. Directional asymmetry falls by nearly half across eras and by approximately seventy percent from its 2014 peak, implying that the market has moved steadily closer to Markov mobility efficiency as it has developed.

Second, the associated return predictability proves largely illusory from an investment perspective. The apparent bottom-decile premium is economically large on paper, particularly during the market's frontier phase, but it is concentrated almost entirely among assets with negligible liquidity, elevated turnover, high delisting risk, and extremely limited portfolio capacity. Two independent arguments dispose of it. Re-forming deciles inside the tradable universe---asking whether the smaller assets an investor could hold earn more---leaves a consolidation-era gradient of $0.01\%$ per week, and a cross-section that is shallowly U-shaped rather than monotone. And even on the construction most favourable to the strategy, a portfolio held to a tenth of its thinnest constituent's daily volume carries a median of \$293,000, on which the net-of-cost return amounts to under four thousand dollars a week. The cryptocurrency size effect is therefore best understood as an illiquidity mirage rather than a scalable investment opportunity: not a premium that vanishes under scrutiny, but one that belongs to assets selected for being small relative to their liquidity, and that cannot be made to carry money.

Third, the convergence of asymmetry is primarily driven by changes in market composition rather than by changes in individual assets. The decomposition introduced in Section~\ref{sec:owncap} shows that the declining component of asymmetry reflects the contraction of entry-driven churn. The frontier market was characterised by a rapid flow of new listings that repeatedly displaced incumbent assets through the relative ranking. As entry slowed relative to the installed base, this source of mechanical mobility weakened, and the market's composition stabilised. We report, but do not lean on, two further pieces of evidence pointing the same way: directional asymmetry is lower in broader market states in weekly data, though breadth and the entry margin cannot be separated once inference is corrected for overlapping windows, and an event study around the spot exchange-traded-fund milestones finds no discontinuity, on a design too low-powered to have found one.

What remains after the contraction of asymmetry is not randomness but structure. The attention-and-survival framework developed in Section~\ref{sec:framework} provides one interpretation of that structure, and the evidence divides it in two. Outside the exit-truncated bottom of the distribution, larger assets exhibit greater positional persistence, and probabilities of disappearance fall with size throughout; formal many-inequality tests do not reject the survival ordering on any universe we examine, and do not reject the persistence ordering on any universe whose deciles are re-formed within it. These patterns do not establish recognition as a uniquely identified causal mechanism. They are consistent with a market in which information acquisition is costly, investor recognition varies systematically across assets, and survival is itself an economic dimension.

The compensation channel is different, and we end on it because it is where the two halves of the paper meet. Surviving small assets do earn higher conditional returns on the capitalisation-reporting population, and the many-inequality tests do not reject that ordering. But in the consolidation era it is rejected once a trading-volume screen is imposed, at the \$10,000 and \$1,000,000 thresholds under either decile construction and marginally short of rejection at \$100,000 ($p=0.053$), and the direct portfolio evidence of Section~\ref{sec:invmargin} agrees: among tradable assets the size gradient in returns is absent. The premium that motivates the framework and the mirage that Section~\ref{sec:payoff} documents are the same object: neglect is compensated precisely where neglect is severe enough to preclude trading, and the compensation disappears at exactly the point where an investor could begin to collect it. A recognition premium that cannot be harvested is not a puzzle for market efficiency. It is a restatement of how illiquid the neglected fringe is.

The broader lesson extends beyond cryptocurrencies. Markets need not be born efficient, and efficiency can emerge as populations stabilise and the disruptive effects of entry diminish. During that process statistical anomalies may persist long after their economic significance has gone, concentrated in a non-investable fringe rather than in the investable core. Distinguishing the two domains is essential, and the distinction is not merely a robustness exercise: as this paper shows, a mechanism can be strongly supported on one and rejected on the other. Population data make that test possible, because they contain the fringe. Samples restricted to tradable assets cannot detect a premium that lives outside them, and samples that pool the two cannot tell which domain a result belongs to.

By combining a mobility-based definition of efficiency with population-level evidence on entry, exit, and survival, the paper offers a new perspective on how asset markets evolve. The cryptocurrency market provides an unusually rich laboratory, but the framework is more general. Whenever assets move through a cross-sectional ranking, questions of mobility, persistence, survival, and efficiency are inherently linked. Understanding how those forces interact may offer new insights into the process through which markets become informationally efficient.

\end{document}